\documentclass[%
 amsmath,amssymb,
 aps, 
 physrev,
]{revtex4-2}

\usepackage{graphicx}
\usepackage{dcolumn}
\usepackage{bm}
\usepackage[mathlines]{lineno}

\begin{document}


\title{\textbf{Effect of Expansion Geometry on Turbulence in Axisymmetric Pipe Flows} 
}%

\author{Jibu Tom Jose}
\author{Gal Friedmann}%
\author{Dvir Feld}
\author{Omri Ram}
 \email{omri.ram@technion.ac.il}
\affiliation{%
 Faculty of Mechanical Engineering, Technion - Israel Institute of Technology, Haifa 3200003, Israel
}%


\begin{abstract}
We investigate the influence of expansion geometry on the flow field and turbulence structure in axisymmetric pipe flows through comparative analysis of abrupt ($90^\circ$) and gradual ($45^\circ$) area expansions with an area ratio of 2.56 at step-height Reynolds numbers of 25000 and 35000. Utilizing refractive index-matched stereo Particle Image Velocimetry, we resolve the three-component velocity fields and extract turbulence statistics with high spatial fidelity. Both configurations exhibit full flow separation, recirculation, and shear layer development; however, the gradual expansion consistently yields elevated turbulence levels, broader shear layers, enhanced Reynolds stress anisotropy, and stronger out-of-plane fluctuations. In contrast, the abrupt expansion generates a secondary vortex that disrupts the return flow, reducing shear layer interaction and turbulent kinetic energy (TKE) production. The governing mechanism is attributed to the geometry-induced modulation of the return flow. In the gradual case, the return flow remains attached to the sloped surface and impinges obliquely on the free-stream, generating a distributed region of high shear and sustained turbulence production leading to intensified TKE and anisotropy in the near-expansion region. The abrupt case confines this interaction, limiting turbulence generation spatially and structurally. These findings reconcile prior observations of increased pressure loss in sloped expansions and reveal the fundamental role of expansion slope in controlling turbulence generation and energy redistribution in separated flows. The observed trends suggest a generalizable mechanism relevant to a broader range of expansion angles and flow conditions.
\end{abstract}

\maketitle

\section{Introduction}
\label{sec:intro}

When a flow encounters a sudden expansion, it typically separates from the surface, forming a shear layer and a recirculation zone. This separation often results in increased drag and energy losses in vehicles, buildings, and fluid systems. In particle-laden flows, the recirculation promotes sedimentation, while in high-speed flows, the resulting low-pressure regions can induce cavitation. Conversely, in some systems, such as mixing chambers and combustors, flow separation can be beneficial, enhancing fuel dispersion and improving combustion efficiency through strong shear. Hence, the flow across a sudden expansion has been studied by many through experimental and numerical tools, focusing on understanding and predicting various parameters such as the reattachment distance  \citep{eaton1980turbulent, otugen1991expansion}, separated flow region dynamics \citep{le1997direct, nadge2014high}, and shear layer development \citep{eaton1982low, scharnowski2014investigation}. As discussed below, due to the technological constraints involved, most studies focused on the case of a Back Facing Step (BFS) with a 90$^\circ$ expansion and primarily examined a two-dimensional geometry. Few studies have investigated axisymmetric area expansions in pipes, and even fewer have addressed cases involving sloped or gradual expansions.

\begin{figure}
    \centering
    \includegraphics[width=0.9\linewidth]{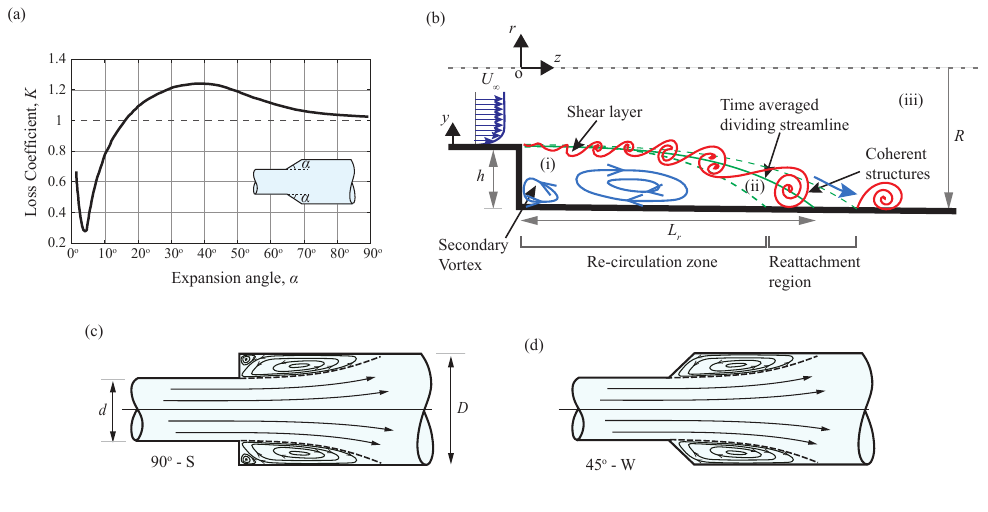}
    \caption{(a) Loss coefficient in an axisymmetric area expansion for various expansion angles, $\alpha$,  based on \cite{White1998}. The loss coefficient, $K$, is calculated by normalizing the measured head loss, $h_m$ with the mean inlet flow speed, $U_m$, and gravity, $g$, ($K=h_m/[U^2_m/(2g)]$). (b) A schematic description of the flow field that develops downstream of a BFS. The main features include a shear layer that separates the incoming flow and a recirculation region. A schematic of the time-averaged flow field in a pipe with axisymmetric area expansion through a (c) 90$^\circ$ step (S) and (d) 45$^\circ$ wedge (W) expansion. }
    \label{fig:bfs_schematic}
\end{figure}

Researchers have extensively explored topics relating to the nature of turbulent pipe flow like laminar-to-turbulent transitions, friction, and the internal structure of the turbulent flow \citep{mckeon2004friction,  marusic2010wall}. These studies emphasize the complexity and diversity of phenomena that exist in turbulent pipe flow. However, most studies on circular pipes have focused on the case of uniform cross-sectional area. A sudden change in the cross-sectional area of the pipe will form an axisymmetric BFS. Early experiments have shown that in cases that involve sudden area expansion at an angle higher than $\sim$16$^\circ$ in a circular pipe, a 90$^\circ$ sudden BFS expansion is preferred as it results lower losses as shown in Fig.~\ref{fig:bfs_schematic}(a) \citep{gibson1910flow, idelchik1986handbook, White1998}. While well known, this phenomenon has never been explained from direct measurements of the flow field. Furthermore, despite being a fundamental problem in fluid mechanics and occurring in many piping systems and devices, experimental data on turbulent flow through an axisymmetric expansion is limited due to technological challenges in imaging flow through the curved walls of a non-uniform cross-section geometry. In particular, obtaining data from imaging-based techniques such as Particle Image Velocimetry (PIV) is challenging due to distortions caused by the curved surfaces of the pipe and strong reflections caused by the intense illumination. However, some studies such as \citet{hammad1999piv} did use PIV for the case of a laminar flow through axisymmetric expansion, but to resolve the optical issues described above, employed the use of refractive index matching technique \citep{amini2012investigation} by using diethylene glycol as the working fluid, which is  $\sim$40 times more viscous than water. Additional investigations using 2D PIV to study axisymmetric sudden expansion in pipe flows include \citet{mak2007near} exploring swirl flow after the expansion, \citet{goharzadeh2009experimental} focusing on laminar flow in annular geometries (index matched with silicone oil), and \citet{hammad1999laminar} examining laminar flow involving non-linear viscoplastic fluids (index matched with diethylene glycol-benzyl alcohol-water mixture).  Other techniques employed to study the problem of an axisymmetrical expansion in a pipe are dye visualization \citep{back1972shear}, surface pressure measurements \citep{devenport1993experimental}, Laser Doppler Velocimetry \citep{morrison1988three,durrett1988radial, stieglmeier1989experimental, gould1990investigation, deotte19913}, and ultrasonic velocity profiler \citep{furuichi2003spatial}. Several theoretical and computational studies have been conducted to address the lack of experimental data on sharp axisymmetric expansion, which leads to flow separation and elucidating the flow characteristics, especially in turbulent flow conditions, \citep[e.g.,][]{teyssandiert1974analysis, cantwell2010transient, javadi2015and, selvam2015localised, lebon2018new}. These studies provide a wealth of information about fluid flow and provide a foundational basis for designing fluid systems. However, because of the turbulent nature of the flow, especially at higher Reynolds numbers, they require the use of various modeling techniques. Therefore, such studies would greatly benefit from more refined and exhaustive experimental data.

 A schematic description of the flow field that develops behind a BFS is shown in Fig.~\ref{fig:bfs_schematic}(b) based on reviews by \citet{bradshaw1972reattachment} and \citet{chen2018review}. The flow in the wake of a BFS can be categorized into three main regions: (i) recirculation region, (ii) shear layer, and (iii) outer region, as shown in Fig.~\ref{fig:bfs_schematic}(b). As the flow separates at the corner of the BFS, it forms a thin shear layer that grows as it progresses downstream, encasing the turbulent eddies generated at the corner. A recirculation is formed with a large clockwise primary vortex and a smaller counterclockwise secondary vortex. The shear layer curves towards the wall and impinges on the wall at the reattachment region. A train of primary vortices has been shown to extend from the step until the reattachment point \citep{scarano1999pattern}. Secondary quasi-streamwise vortices have shown to form between the primary vortices \citep{bernal1986streamwise,bell1992measurements,agarwal2023pressure}. 

Earlier studies have characterized various flow features associated with fully separated flow caused by an abrupt area increase. These have shown that the mean reattachment length, $L_r$ is about 5-8 times the step height depending on Reynolds number and expansion geometry \citep{kim1978investigation, otugen1991expansion, kostas2002particle}. However, it has been shown that $L_r$ varies significantly and is typically characterized by an unsteady oscillatory motion with a flapping frequency that effectively changes the size of the recirculation region \citep{eaton1982low, bhattacharjee1986modification, ma2017analysis}. Numerous studies have examined the unsteady nature of the flow using various time-resolved measurements, including hot-wires \citep{spazzini2001unsteady}, high-speed PIV \citep{sampath2014proper, ma2022investigation}, and numerical simulations \citep{schafer2009dynamics}. Additional studies examined the characteristics of the shear layer above the separated region \citep{browand1966experimental,winant1974vortex, troutt1984organized}, and the turbulence properties associated with both the shear layer and the recirculation zone \citep{kasagi1995three, fessler1999turbulence, piirto2003measuring}. As the shed vortices affect heat, mass, and momentum transfer in the wake of separated flows \citep{hussain1986coherent}, their advection downstream and the subsequent formation of coherent structures were also extensively studied \citep[e.g.,][]{lee2004three, wang2019experimental}.

As mentioned above, a survey of the literature discussed above reveals that almost all studies have focused on the case of a BFS expansion. There are only a few studies that considered the effect of more gradual expansion, and most of them relied on numerical methods \citep[e.g.,][]{selvam2015localised, ahmadpour2016numerical, choi2016numerical, danane2020effect}. This study compares the case of a sharp, 90$^\circ$ expansion, and a sloped, 45$^\circ$ expansion in a pipe. A schematic of the two cases and the mean flow fields in the separation regions is depicted in Fig.~\ref{fig:bfs_schematic}(c) and (d). In both cases, the flow separates at the corner of the area expansion. However, while it is well known that the 45$^\circ$ expansion case creates higher pressure losses than in the 90$^\circ$ expansion, the specific flow dynamics and governing mechanisms remain unclear. Additionally, the experimental data on turbulent flow axisymmetric expansion are limited, particularly in comparing the flow characteristics and turbulence statistics between abrupt and gradual expansion cases. This study employs refractive index matching, which allows for unobstructed optical access across the expansion section, facilitating stereo PIV measurements and enabling the analysis and comparison of outcomes in both scenarios to discern the essential similarities and differences in the flow fields and turbulence properties downstream of the expansion. 

The subsequent sections are structured as follows. First, we provide an overview of the experimental setup and techniques, including the refractive index-matched tunnel, PIV configurations, and data processing methods. The following results section begins by presenting the properties of the flow field upstream of the expansion followed by detailed measurements of the mean flow fields and turbulence statistics, focusing on the comparison between step and gradual expansion cases. Additionally, Reynolds stress anisotropy analysis is used to highlight the variance in turbulence structure and elucidate the governing mechanism behind its dependence on expansion geometry.

\section{Experimental Methodology and Data Analysis}
\label{sec:expt}

\subsection{Experimental setup}
\label{ssec:piv}

\begin{figure}
    \centering
    \includegraphics[width=1\linewidth]{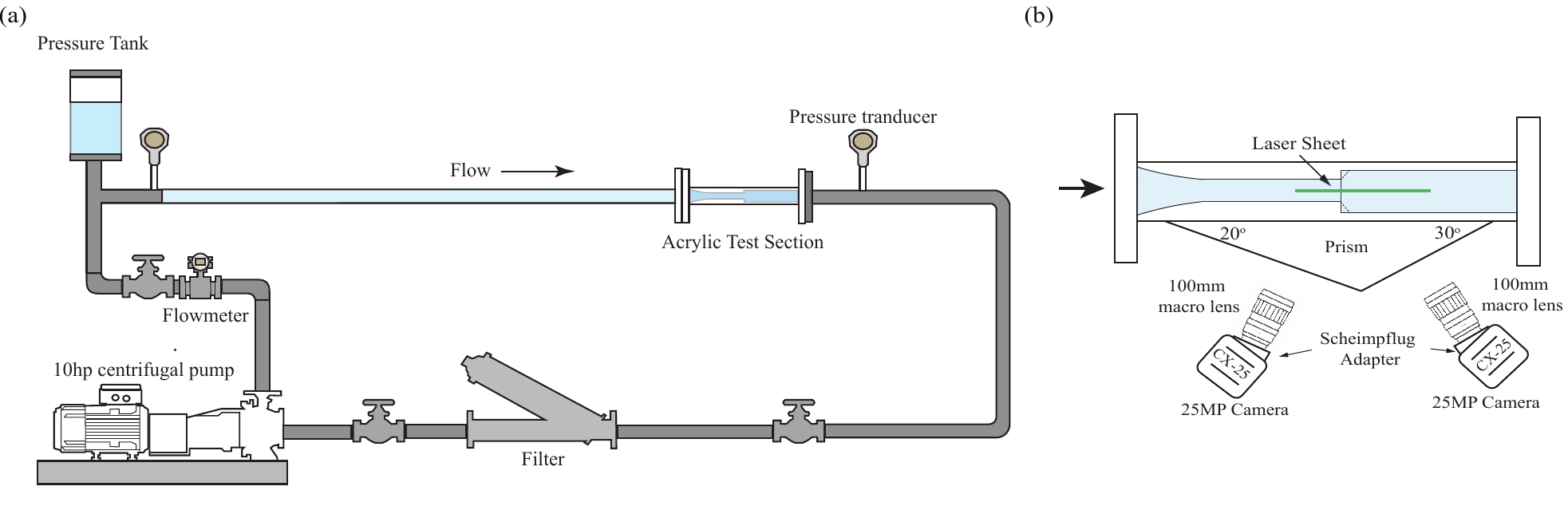}
    \caption{(a) Schematic of the refractive index matched water tunnel highlighting the  major features; (b) Stereo PIV configuration using 2 CX-25 PIV cameras}
    \label{fig:channel_PIV}
\end{figure}

The experiments are conducted in the refractive index-matched water tunnel facility at the Transient Fluid Mechanics Laboratory at the Technion. Fig.~\ref{fig:channel_PIV} provides a schematic description of the experimental setup. The loop is fitted with a 10 hp centrifugal pump (SAER NCBXSD80-200/B) that can circulate up to 130m$^3$/h of water, enabling a maximum mean flow of 10m/s at the test section. The pump is controlled using a variable frequency drive (Nastec Vasco V418) to hold a steady flow rate during experiments. The flow rate in the loop is limited to about 6m/s to prevent vibration during experiments occurring at high flow rates. The flow rate is measured using an electromagnetic flowmeter (Flowtech kf710-80) with a mass flow accuracy of $\pm0.5\%$ of the measured flow rate. A clear pressure tank (SR-TEK 1000-Cl-CT) is connected at the top of the loop, which serves three main purposes. (i) The tank is kept half full, and the volume of gas is pressurized using an external nitrogen supply. This is used to set the pressure within the loop. Nitrogen is used to prevent oxidization of the working fluid in the loop. (ii) The tank is fitted with a safety release to restrict the maximum pressure to prevent over-pressurizing the system, which is mainly constructed from plastic materials. (iii) Excess gas trapped in the system during filling is collected in the tank and during the de-aerating procedure. De-aeration of the system is performed to reduce the likelihood of bubble formation at higher flow rates by connecting the pressure tank to a vacuum pump (Bosch R5) for an extended duration. Two pressure transducers (WIKA UPT-20) were installed on the loop to monitor system pressure and ensure safe operation within the tunnel’s structural limits. The loop is fitted with a 2000mm long, 40mm diameter clear acrylic straight pipe upstream of the test section entrance, providing unobstructed flow for 50 pipe diameters, ensuring a fully developed flow at the entrance of the test section. A 300mm-long replaceable test section is machined out of clear acrylic (PMMA), providing optical access and enabling the study of internal flow with various geometries. Each internal geometry requires the fabrication of a new test section. A concentrated 62-63\% aqueous solution of sodium iodide (NaI) with a refractive index of 1.488, which matches that of the acrylic, is utilized to facilitate undisturbed imaging of the internal flow through the curved walls of the test section and the sharp geometry change. The NaI solution has a kinematic viscosity of $1.1\times 10^{-6}$m$^2$/s at $20^\circ$C and a specific gravity of 1.820 \citep{bai2014refractive}. The outer walls of the test sections are machined flat, and acrylic prisms are installed to ensure the cameras are positioned normal to the test section, keeping the lenses perpendicular, preventing reflections, and reducing distortions.

This study uses two different geometries, as shown in Figs.~\ref{fig:bfs_schematic}(c) and (d). The first geometry is a sharp area expansion, where the pipe diameter gradually reduces to a diameter of 25mm at the entrance to the test section allowing the flow to fully develop for a distance of $\sim$10 diameters and then sharply expands through a BFS into the straight pipe with a diameter of 40mm, resulting in a step height, $h$, of 7.5mm and an area expansion ratio of 2.56. The second geometry is similar, but the expansion occurs more gradually through a 45$^\circ$ slope. The experiments have been conducted with step-height-based Reynolds numbers ($Re_h=U_m h/\nu$) of 25000 and 35000 for both geometries, using mean inlet velocities, $U_m$, of 3.6 m/s and 5.1 m/s, respectively. The velocities are chosen based on the limits of the experimental setup to prevent pump stall or cavitation. The 90$^\circ$ step case is denoted as S-25 and S-35, corresponding to $Re_h$=25000 and $Re_h$=35000, respectively. Similarly, the 45$^\circ$ expansion is denoted as W-25 and W-35 in subsequent sections. 
 
We use stereo-PIV to record the flow field downstream of the area expansion, and focuses on acquiring the turbulent flow field properties downstream of the step. A schematic of the stereo PIV system is provided in Fig.~\ref{fig:channel_PIV}(b). The stereo-PIV experiments are performed using two 25 Megapixel cameras (LaVision Imager CX-25) fitted with 100mm macro lenses, $f/5.6$, using a resolution of 5993$\times$3035 pixels with a magnification of 55.4 pixel/mm.  Scheimpflug adapters are used to focus the cameras on the laser sheet. Each test setup, defined by its geometry and Reynolds number, has been documented for 200 seconds, providing 3000 instantaneous flow field realizations, which are used to calculate the ensemble-averaged flow fields. A double pulse Nd:YAG laser (Quantel EverGreen 532-200) is used to illuminate the flow field with a 1mm thick laser sheet. Silver-coated hollow glass spheres with a diameter of  13$\mu$m  are used as tracer particles (Stokes number = 8$\times 10^{-4} \ll 1$). A summary of the PIV imaging setup properties is given in Table~\ref{table:PIV}. PIV calibration is performed by replacing the test section with an identical rectangular tank with an open top filled with NaI. A three-dimensional calibration plate (LaVision 058-5-1) is placed at the location of the laser and calibration images are acquired.

\begin{table}
    \centering
    \begin{tabular}{ccc}
    \hline
        \textbf{Experiment} & \textbf{Stereo PIV} \\
          \hline
        \textbf{Field of View} (mm) & 107$\times$54 \\
        \textbf{Frequency} (Hz) & 15 \\
        \textbf{Vector Resolution} ($\mu$m) & 217   \\ 
        \textbf{Sum-of-Correlation} ($\mu$m) & 54   \\ 
        \textbf{Snapshots} &  3000 \\ 
        \textbf{Run Time} (s)  & 200 \\ 
        \textbf{Cameras} & 2$\times$ Imager CX-25 \\
        \textbf{Tracer Particles} & 13$\mu$m silver-coated hollow glass spheres \\
  \hline
    \end{tabular}
    \caption{Parameters of the stereo PIV imaging setup used in this study.}
  \label{table:PIV}
\end{table}

\subsection{Data Analysis}
\label{ssec:analysis}

The flow field is obtained using LaVision DaVis\texttrademark 11 PIV software. Planar self-calibration is performed to improve the accuracy of PIV measurements. The images are pre-processed using sliding background subtraction to remove minor reflections and enhance the raw images. Coarse calibration is followed by a self-calibration procedure to improve the accuracy of the stereo PIV data. The cross-correlation-based PIV analysis is performed with a final interrogation window size of 24$\times$24 pixels with 50\% overlap. Additionally, the mean flow field is calculated from the Stereo-PIV using sum-of-correlation analysis \cite{meinhart2000piv} using 3000 images, which enables a final interrogation window size of 12$\times$12 pixels with 75\% overlap. The analysis results in a final vector resolution of 217$\mu$m for standard cross-correlation analysis and 54$\mu$m for the sum-of-correlation analysis. The associated uncertainties are discussed in Appendix~\ref{app:Uncertainty}. The Kolmogorov length scale for the fully developed inlet flow is estimated using $\epsilon \sim~u_{rms}'^3/l$, where the integral length scale $l$ is calculated from 1D correlation of streamwise velocity at $z^*$=1.5 upstream of the expansion. The Kolmogorov length scale is then calculated as $\eta = (\nu^3/\epsilon)^{1/4}$ and about 43 $\mu$m. This gives the vector resolution of $\sim$5$\eta$ for both the Reynolds numbers. In the case of sum-of-correlation, the resolution is $\sim$1.26$\eta$.

In all subsequent sections, $r$ is measured from the center line of the pipe, and $y$ is measured from the wall. The streamwise coordinate, $z$, is measured in relation to the onset of the expansion where $z$=0 (see Fig.~\ref{fig:bfs_schematic}(b)). The normalized quantities $r/h$ and $z/h$ are represented as $r^*$ and $z^*$ respectively. Using these definitions, the step is located at $ r^*$=1.67, and the wall of the expanded section downstream of the step is at $ r^*$=2.67. The fluctuating velocity components, $u_i'$, are calculated as $u_i'=U_i-\langle U_i \rangle $, where $U_i$ is the instantaneous velocity at a point, $\langle \cdot  \rangle $, is the ensemble average and $i$=1-3 corresponds to the streamwise, radial and azimuthal direction denoted as $u_z,u_r$ and $u_\theta$, respectively.

\section{Results}
\label{sec:results}

\subsection{Inlet flow conditions}
\label{ssec:inlet}

\begin{figure}
    \centering
    \includegraphics[width=0.95\linewidth]{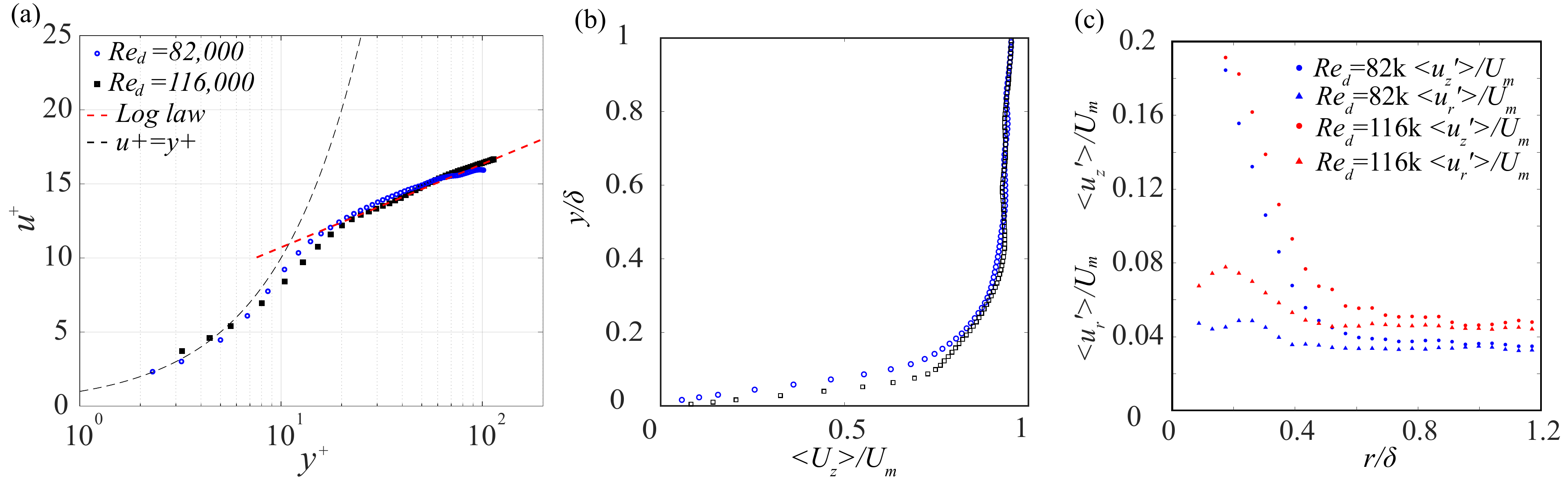}
    \caption{(a) Velocity profile upstream of the step at both the Reynolds numbers. The log-law is plotted in red dashed line. (b) Mean velocity profile upstream of the expansion and (c) Mean turbulence intensity upstream of the step for $Re_d$=82000 and $Re_d$=116000.}
    \label{fig:bl_profile}
\end{figure}

Fig.~\ref{fig:bl_profile} shows the mean flow profile upstream of the expansion region for the two flow rates used in this study. We have performed experiments using two flow rates. The first flow rate of $Q$=1.77$\times10^{-3}$ m$^3$/s resulting in an inlet diameter-based Reynolds number, $Re_d=U_md/\nu$, of 82000  and a second flow rate of $Q$=2.5$\times10^{-3}$ m$^3$/s for which $Re_d$ is 116000. The turbulence quantities are expressed in the wall coordinates denoted by a '+' superscript. Using the high spatial resolution sum-of-correlation PIV data, the normalized velocity is calculated and plotted in Fig.~\ref{fig:bl_profile}(a). The dashed red line shows a fit to the log-law, $ u^+ = (1/\kappa) \ln(y^+)+B$, where  $y^+=yu_\tau /\nu$,  $u^+ = \langle U_z \rangle /u_\tau$, $\kappa$=0.41 and $B$=5.1 from which we extracted the friction velocity,  $u_\tau$. The velocities in Fig.~\ref{fig:bl_profile}(a) show good agreement with the log-law, indicating a fully developed turbulent flow upstream of the expansion. Based on $u_\tau$, $Re_\tau=u_\tau R/ \nu \approx $ 2500  and $Re_\tau \approx $ 3300 corresponding to $Re_d$=82000 and $Re_d$=116000, respectively. Additionally, Fig.~\ref{fig:bl_profile}(b) plots the mean velocity profile  showing the effects of $Re_d$.  Fig.~\ref{fig:bl_profile}(c) shows the turbulence intensities far upstream of the expansion region and is less than 5\% in the free-stream.

\subsection{Convergence and axisymmetry of the flow}
\label{ssec:axis}

The velocity profiles show that the flow field has converged based on the 3000 realizations including the region inside the recirculation region. As expected, we have measured $\langle u_i' \rangle$ to be zero for all the data sets ($\mathcal{O}(10^{-14})$). The mean velocity profiles are almost symmetrical for the upper and lower halves, confirming that the mean flow is axisymmetric. Furthermore, we have measured the out-of-plane velocity  $\langle U_\theta \rangle $ to be about $0.05\langle U_m \rangle $ in the shear layer and negligible in the recirculation region (less than $0.01\langle U_m \rangle $). The out-of-plane Reynolds stress, $\langle u_\theta'u_\theta'\rangle$ is comparable to the wall-normal Reynolds stress $\langle u_r'u_r' \rangle $ (presented in section~\ref{ssec:TKE_re}). Additionally, $\langle u_z'u_\theta' \rangle $ and $\langle u_r'u_\theta' \rangle $ are about one order of magnitude smaller than $\langle u_z'u_r' \rangle $. This further supports that the mean flow is axisymmetric.We have evaluated the statistical independence of the velocity fields  by computing the autocorrelation of streamwise velocity fluctuations $u'(t)$  at representative locations along the centerline, shear layer, and recirculation region, following the procedure outlined by \citet{saarenrinne2001experiences}. The de-correlation time $ \tau_c $ is estimated as the integral of the autocorrelation function up to the first lag at which it drops below 0.05, with the maximum observed value of $ \tau_c \approx $ 0.033~s. Since the sampling interval $\Delta t$ = 0.0667~s exceeds $\tau_c$, all 3000 snapshots are considered statistically independent and suitable for ensemble averaging. We have found that the uncertainty in mean velocity components is less than 0.5\% of $U_m$ and less than 0.2\% of $U_m^2$ for mean Reynolds stress terms. More details on the uncertainty analysis of PIV data are provided in Appendix  ~\ref{app:Uncertainty}.

\subsection{Mean flow fields}
\label{ssec:meanflow}

\begin{figure}
    \centering
    \includegraphics[width=0.95\linewidth]{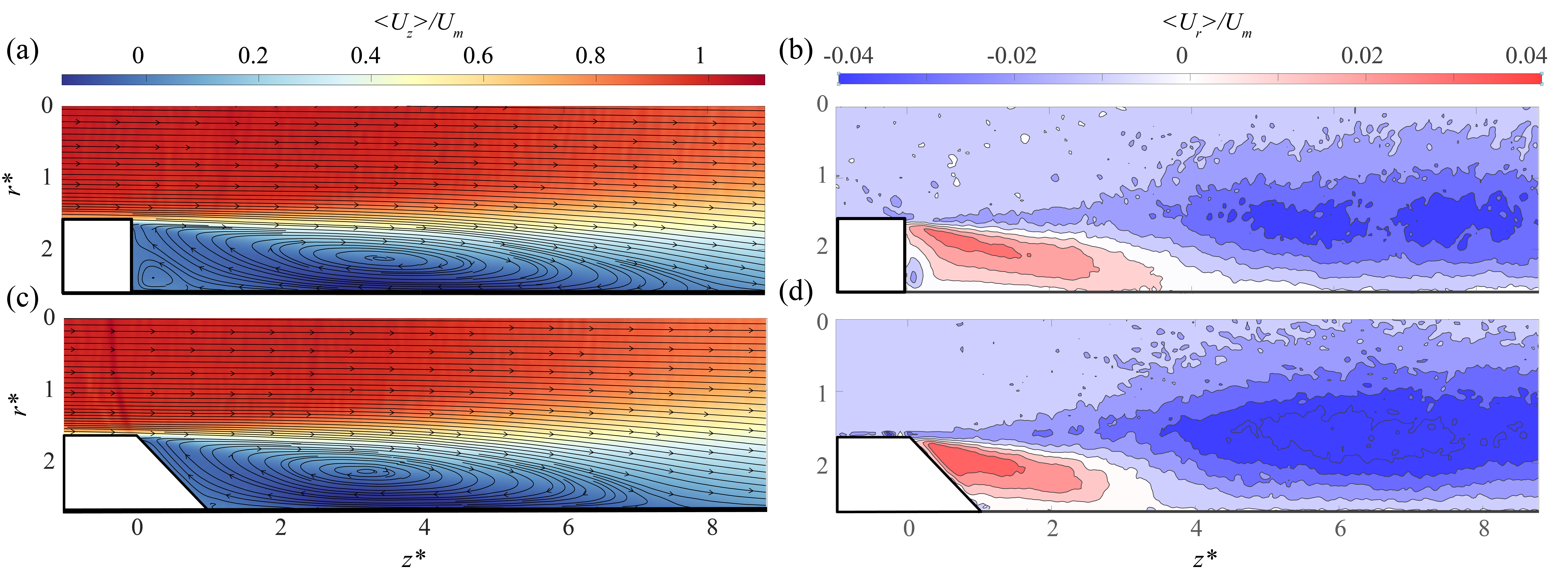}
    \caption{Maps of normalized streamwise velocity plots for (a) S-25 and (c) W-25 overlaid by the flow streamlines. Contours of normalized wall-normal velocity plots for (b) S-25 and (d) W-25. All the plots are for the bottom half of the test section.}
    \label{fig:velocitymaps}
\end{figure}

Fig.~\ref{fig:velocitymaps} depicts the mean flow field obtained by ensemble averaging 3000 realizations from experiments performed at $Re_h$=25000 for both step (S-25) and wedge (W-25) cases. The mean streamwise flow is plotted in Figs.~\ref{fig:velocitymaps}(a) and (c), overlaid by the flow streamlines, and the wall-normal velocity is plotted in Figs.~\ref{fig:velocitymaps}(b) and (d). Although not shown, the out-of-plane velocity, $\langle U_\theta \rangle$ is about 0.05$\langle U_z \rangle$ in the shear layer and negligible everywhere else. A primary recirculating vortex forms in both cases, however a counter-rotating, secondary vortex appears only in the step cases. For the step case, the location of the primary vortex core is at $z^*$=3.5 and $y$=0.5$h$, and the secondary vortex core is found to be at $z^*$=0.3 and $y$=0.2$h$. The core of the primary vortex in the wedge case is slightly shifted upstream at $z^*$=3.2 and $y$=0.5$h$. The dynamics of the recirculation region in the decelerating core region are governed primarily by adverse pressure gradients. Hence, as the flow goes across the area expansion, it is to be expected that both cases behave similarly. In $\langle U_z \rangle$, the significant changes appear to be limited to the region in the immediate vicinity of the corner. However, distinct differences are seen in the wall-normal velocity, $\langle U_r \rangle$. A clear $\langle U_r \rangle$=0 line separates two regions in the recirculation region, which traverse from the corner, through the core, and to the wall. Comparing the regions of low wall-normal velocity shows that they exhibit different morphologies. In the wedge case, this region appears more confined and closer to the core. However, in the step case, the low-wall normal velocity extends beyond the core and up to $z^*$=6. This difference stems from the core of the primary vortex being inclined in the step case, while in the wedge case, it is horizontal. Furthermore, this coincides with the behavior of $\langle U_r \rangle$, close to the wall, in the backflow region upstream of the core. In the wedge case, there exists a significant region with very low $\langle U_r \rangle$, which suggests that the backflow here is well organized and the streamlines are parallel to the wall. In comparison, in the step case, the region where the streamlines are parallel to the wall is more limited, with an upflow starting well before the secondary recirculating vortex. Additionally, the presence of the secondary vortex creates a second low $\langle U_r \rangle$ region that corresponds to its core. As the flow returns upstream, the secondary vortex forces the flow to separate from the wall, flow around it, and reattach to the vertical face of the step shortly before it impinges on the incoming flow. Conversely, the flow in the wedge case does not separate from the wall and continues along the inclined face as it flows upstream to the corner. A pronounced difference is seen in $\langle U_r \rangle$ in the shear layer region, where the wall-normal velocity increases rapidly for the wedge case and reaches significantly higher levels. Furthermore, the shear layer expands faster upwards into the free-stream, an important feature that we discuss in detail in later sections. We have observed similar trends as those presented above occurring in S-35 and W-35 cases. 

\begin{figure}
    \centering
    \includegraphics[width=0.95\linewidth]{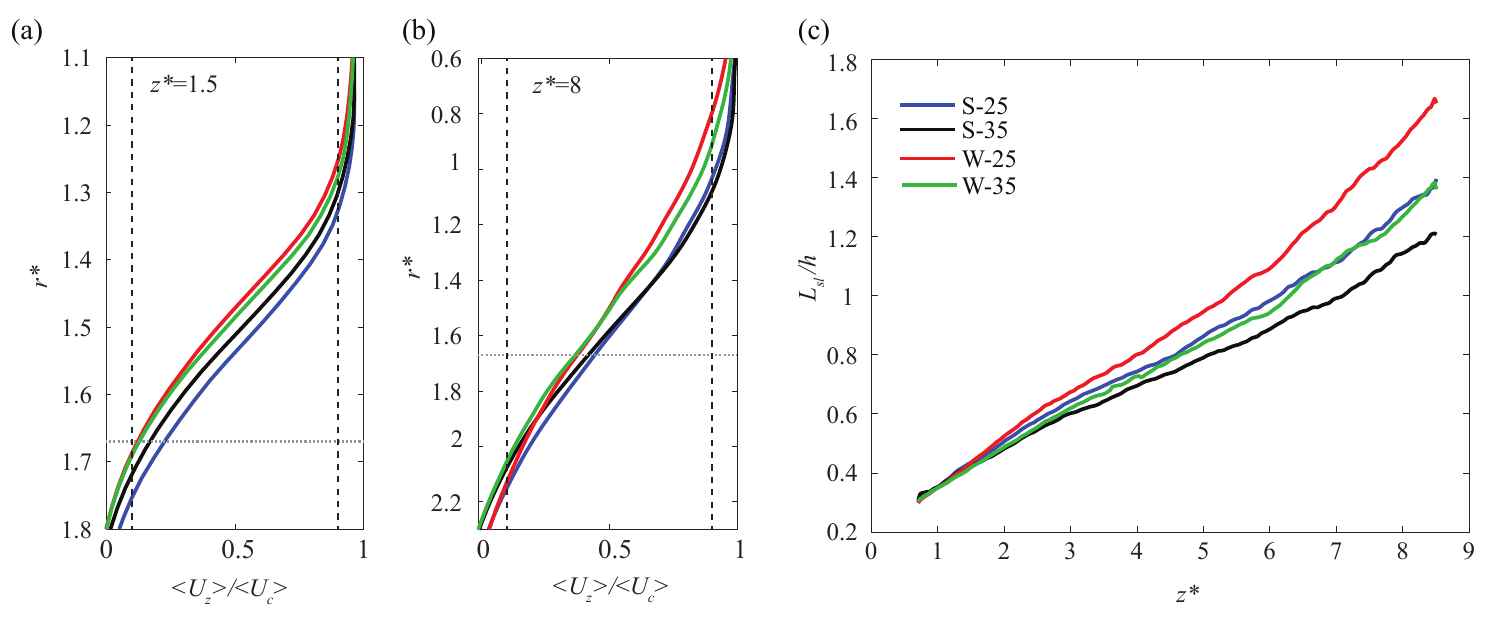}
    \caption{ Mean streamwise velocity in the profiles in the shear layer at (a) $z^*$=1.5 and (b) $z^*$=8 normalized by the center-line velocity. The dotted horizontal line shows the location of the step and the two dashed lines correspond to $0.1\langle U_{z,c}\left( z \right) \rangle $ and $0.9\langle U_{z,c}\left( z \right) \rangle $.  (c) Shear layer thickness $L_{sl}/h$ calculated between $0.1\langle U_{z,c}\left( z \right) \rangle $ and $0.9\langle U_{z,c}\left( z \right) \rangle$.}
    \label{fig:shearlayer}
\end{figure}

Apart from the flow properties near the wall, the region of the shear layer plays a significant role in the evolution of the flow field downstream of the expansion. Figs.~\ref{fig:shearlayer}(a) and (b) plots the mean streamwise velocity profile within the shear layer locations at $z^*$=1.5 and $z^*$=8, respectively. The velocity profiles are normalized for comparison by their respective center-line velocity $\langle U_{z,c} \left( z \right) \rangle $.  A horizontal dotted line marks the step height, and a vertical dashed line represents the 0.1 and 0.9 $\langle U_{z,c}\left( z \right) \rangle $ locations. The shear layer thickness, $L_{sl}$ has been evaluated by previous studies as a means of comparing different mixing layers. Here we use the conservative estimate of the shear layer thickness, similarly to \citet{BrownRoshko1974}. Comparing Figs.~\ref{fig:shearlayer}(a) and (b) show that the shear layer develops in the wedge expands faster and becomes significantly wider than the step cases. Fig.~\ref{fig:shearlayer}(c) plots the width of the 0.1 to 0.9$\langle U_{z,c}\left( z \right) \rangle $ region normalized by the step height, $L_{sl}/h$,  showing that this effect depends both on geometry and Reynolds number, however for both Reynolds numbers we have tested the wedge case shear layer widens faster. Additionally, these plots show that the shear layer appears to be pushed further into the free-stream in the wedge experiments. This effect can also be seen in Fig.~\ref{fig:velocitymaps}(d), by examining the rate by which  $\langle U_r \rangle$ increases downstream of the corner. In both geometries, $L_{sl}$ increases faster in the low Reynold numbers cases where, for W-25, it reaches the center line first, around $z^*$=6, which is upstream of the flow reattachment. The effects of the Reynolds number can be explained by the difference in momentum of the incoming flow. By comparing the mean velocity fields, the maximum velocity deficit in the shear layer $\langle U_m \rangle$ is about 0.15$U_m$ in the wedge cases, compared to step cases. The following sections show that the difference in the observed trends caused by geometry stems from increased levels of turbulent kinetic energy and Reynolds stresses caused by an interaction between the return flow in the recirculation region and the incoming free-stream close to the corner.

\subsection{Turbulent Kinetic Energy and Reynolds stresses}
\label{ssec:TKE_re}
\begin{figure}
    \centering
    \includegraphics[width=0.95\linewidth]{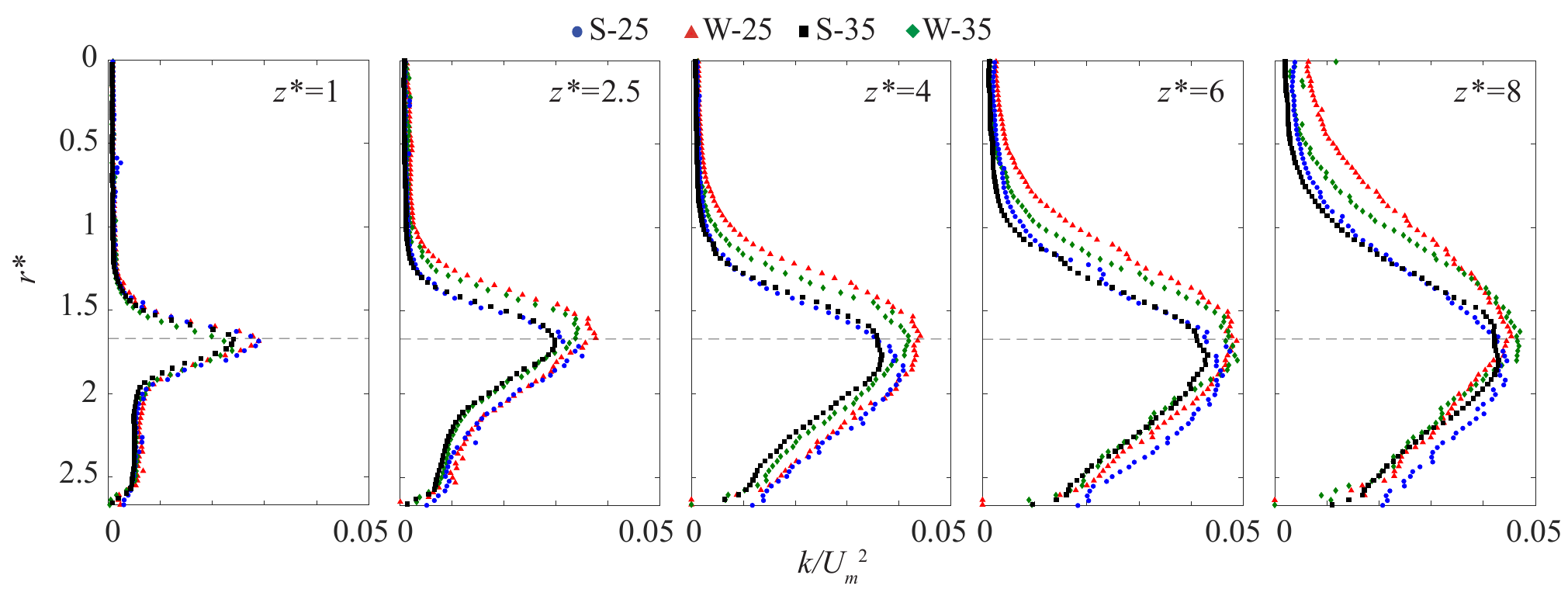}
    \caption{Normalized mean turbulent kinetic energy, $k/U_m^2$,  at $z^*$=1, 2.5, 4, 6 and 8 for S-25, S-35, W-25 and W-35 cases. The dashed horizontal line shows the location of the step. }
    \label{fig:TKE}
\end{figure}

\begin{figure}
    \centering
    \includegraphics[width=0.95\linewidth]{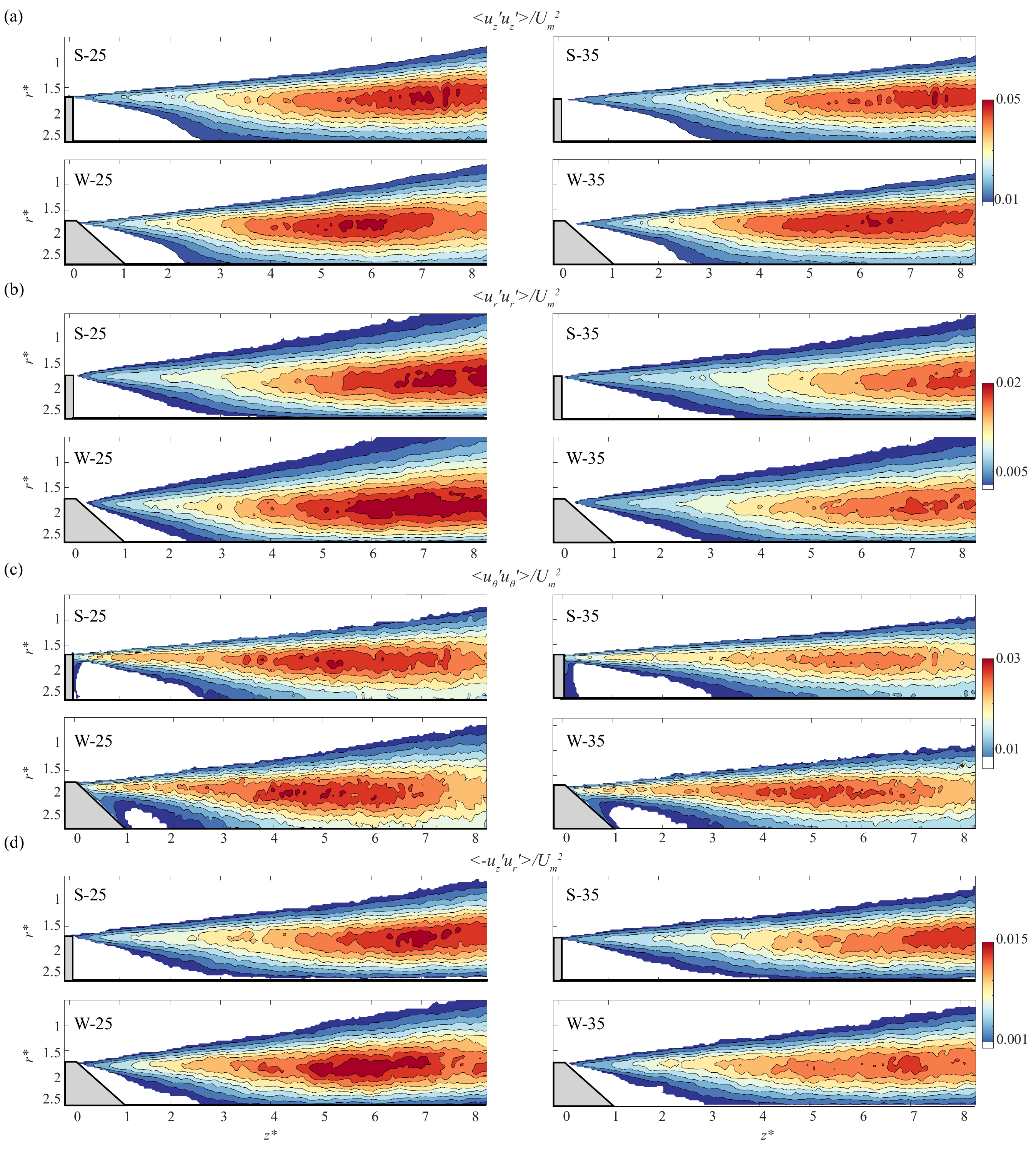}
    \caption{Contour plots of normalized Reynolds stress terms (a) $\langle u_z'u_z' \rangle $, (b) $\langle u_r'u_r' \rangle $, (c) $\langle u_\theta' u_\theta' \rangle $,  and (d) $\langle u_z'u_r' \rangle $ for S-25, S-35, W-25 and W-35 cases. Values lower than lower limits of the colorbar are set to white for easier interpretation}
    \label{fig:reynoldstress}
\end{figure}

Fig.~\ref{fig:TKE} shows the mean turbulent kinetic energy (TKE) profiles normalized by the inlet mean velocity, $k /U_m^2$, at different downstream locations. Fig.~\ref{fig:reynoldstress} presents maps of the normalized mean Reynolds stresses, $\langle u_i'u_j' \rangle/U_m^2$, downstream of the step for all four cases. The turbulent kinetic energy is calculated from experiments as $k=\langle u_i'u_i' \rangle/2 $, and the Reynolds stress as $\tau_{ij}= u_i'u_j'$. The overall behavior of the TKE agrees well with the trends observed in the shear layer. Fig.~\ref{fig:TKE} shows a characteristic progression in $k$ where \mbox{W-25 $>$ W-35 $>$ S-25 $>$ S-35}, with the bulk of the TKE concentrated in the shear layer. Corresponding to the shear layer evolution towards the wall, the peak of the TKE broadens, and the locations of maximum TKE  in Fig.~\ref{fig:TKE} shifts downward, moving away from the center-line. The peak of the TKE appears to be higher in magnitude in the wedge cases before reattachment occurs. Downstream of the reattachment, the region located below the peak in TKE becomes similar for all cases, but the region above the peak does not. The wedge cases consistently show higher levels of TKE from the moment the flow separates, implying that the wedge geometry induces more significant mixing. This feature becomes more pronounced as the flow advances downstream. The relationship between the TKE distribution and shear layer broadening is clearly illustrated by the results of S-25 and W-35, which coincidentally overlap in both (see Figs.~\ref{fig:shearlayer}(c) and~\ref{fig:TKE}). This agreement reinforces our confidence in the measured results, as it demonstrates a clear connection between the rate of shear layer broadening and the generation of TKE. 

We have found that in the bulk of the recirculation region and shear layer, the relative average fraction of $u_z', u_r'$, and $u_\theta'$ constitute 60-65\%, 30-35\%, and 0-5\% of TKE, respectively. However, close to the step corner, the contribution of $u_\theta'$ is far larger and found to be as high as 80\%. This is driven by the interaction between the low-momentum return flow from the recirculation region and the high-momentum streamwise flow above the step. The return flow is unable to penetrate the incoming flow, which results in a strong out-of-plane flow. As the flow progresses downstream, the momentum in the shear layer reduces, and the TKE increases; this is seen throughout our measured domain, which extends to $z^*$=8.5.  

Fig.~\ref{fig:reynoldstress} compares the mean Reynolds stress terms, $\langle \tau_{ij} \rangle$, for all four cases. The results consistently show higher $\langle \tau_{ij} \rangle$ in the wedge cases compared to step cases for both $Re_h$, which aligns with the previously discussed TKE trends and indicates greater turbulent mixing in the case of gradual expansion also evident by higher $- \langle u_z'u_r' \rangle$ (Fig.~\ref{fig:reynoldstress}(d)). Additionally, the location of the peaks of all $\langle \tau_{ij} \rangle$ are further downstream in the step cases. Moreover, all the peaks shift downstream with increasing $Re_h$. The highest magnitude of the normal stress is observed in the streamwise direction, $\langle u_z'u_z' \rangle$, as seen in Fig.~\ref{fig:reynoldstress}(a). The trends of $\langle u_\theta' u_\theta' \rangle $ are different from the rest, with its core appearing further upstream in both step and wedge cases as evident from Fig.~\ref{fig:reynoldstress}(c). The high magnitude of $\langle u_\theta' u_\theta' \rangle $ is consistent with the results reported by \cite{tinney2006low}. Furthermore, as discussed above, we have found that close to the corner ($z^*<$ 1), $\langle u_\theta' u_\theta' \rangle $  is significantly higher, a feature that has not been captured by \cite{tinney2006low} presumably, due to the lack of spatial resolution. A notable feature of the wedge cases is the existence of non-negligible $\langle u_\theta' u_\theta' \rangle $  in the regions of backflow as it approaches back towards the corner. In this region, there is also a considerable out-of-plane velocity component due to the effect of the impinging return flow.

\subsection{Turbulent kinetic energy budget}
\label{ssec:turb_stat}

To study the differences between the wedge and step geometries further, we have calculated the production, convection, dissipation, and diffusion of TKE within the flow. This allows us to qualitatively quantify how energy is generated, transported, and dissipated and provide insight into turbulence dynamics and flow behavior. The TKE budget equation can be expressed as, 
\begin{equation}
\frac{\partial k}{\partial t} + \mathcal{C} = \mathcal{P}_k - \epsilon + \mathcal{D}
\end{equation}

where, the convection term, $\mathcal{C}$, is
\begin{equation}
  \mathcal{C}= U_j \frac{\partial k}{\partial x_j},
    \label{eq:C}
\end{equation}

the production, $\mathcal{P}_k$, is 
\begin{equation}
\mathcal{P}_k = - \langle u_i' u_j' \rangle \frac{\partial U_i}{\partial x_j},
\label{eq:P}
\end{equation}

the dissipation term, $\epsilon$, is
\begin{equation}
    \epsilon = \nu \langle \frac{\partial u_i'}{\partial x_j} \frac{\partial u_i'}{\partial x_j} \rangle,
\label{eq:ep}
\end{equation}

and the diffusion, $\mathcal{D}$, term is
\begin{equation}
\mathcal{D} = \mathcal{D}_v  + \mathcal{D}_t + \mathcal{D}_p
\label{eq:D},
\end{equation}

where, $\mathcal{D}_v$, $\mathcal{D}_t$ and $\mathcal{D}_p$ denote viscous, turbulent and pressure diffusion respectively, and are defined as, 

\begin{equation}
    \mathcal{D}_v = \nu \frac{\partial^2 k}{\partial x_j \partial x_j},     \mathcal{D}_t = - \frac{\partial}{\partial x_j} \left( \frac{1}{2} \langle u_i' u_i' u_j' \rangle \right),     \mathcal{D}_p = - \frac{\partial}{\partial x_j} \left( \frac{\langle p' u_j' \rangle}{\rho} \right). 
\end{equation}

\begin{figure}
    \centering
    \includegraphics[width=0.9\linewidth]{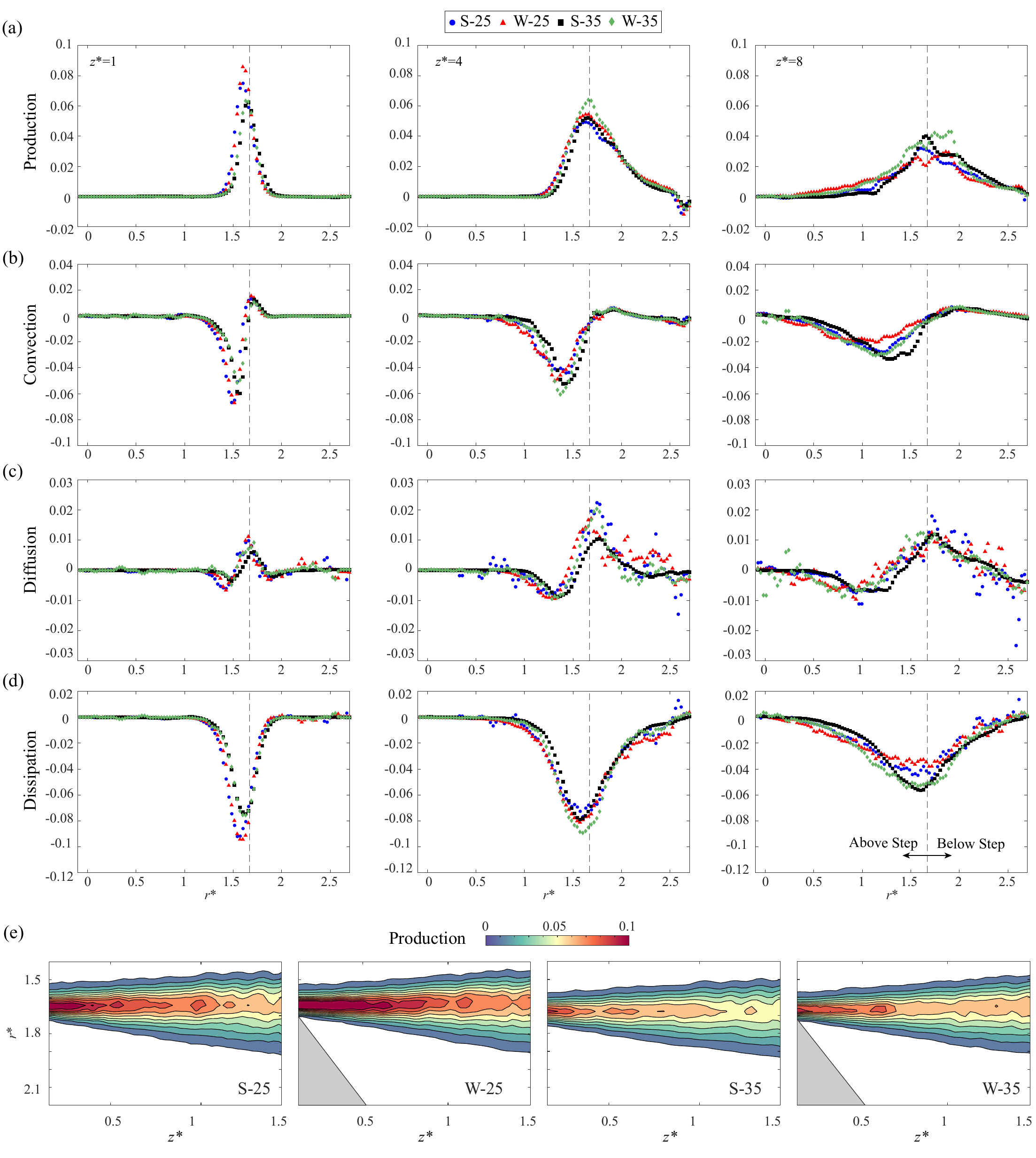}
    \caption{Comparison of turbulent transport terms at various downstream locations for S-25,W-25, S-35 and W-35 cases; (a) Convection; (b) Production ; (c) Diffusion; (d) Dissipation. Left column: $z^*$=1; Middle Column: $z^*$=4; Right column: $z^*$=8. The dashed vertical line is the location of the step. The step extends from $r^*$=1.67 to $r^*$=2.67 (wall). (e) Magnified map of TKE production in the immediate vicinity of the expansion at $0.1<z^*<1.5$ and $1.4<z^*<2.2$ for all four cases.}
    \label{fig:turbsw_all}
\end{figure}

Fig.~\ref{fig:turbsw_all} compares $\mathcal{P}_k$, $\mathcal{C}$, $\mathcal{D}$, and $\epsilon$ at three $z^*$ between all cases. All the quantities are normalized by their respective $U_m^3/h$, and the step height is marked by a dashed vertical line at $r^*$=1.67. The turbulence statistics are calculated from the stereo PIV and hence can not be used to estimate the pressure diffusion, $\mathcal{D}_p$, and the temporal TKE, $\partial k / \partial t$, terms.  The out-of-plane velocity components are included in TKE estimates by using the axisymmetric assumption. The impact of neglecting out-of-plane gradients in the PIV data for BFS has been discussed in detail in \cite{piirto2003measuring}, showing their effects to be negligible. Comparing the turbulent production terms show that it is primarily driven by $\left\langle u_r' u_z' \right\rangle (\partial U_z / \partial r)$. In the case of diffusion,  $\mathcal{D}_t \gg \mathcal{D}_v$ for all cases. The dissipation, $\epsilon$, is calculated as a residual due to insufficient spatial resolution and high sensitivity to noise in higher-order derivatives \citep{kasagi1995three, jovic1996experimental}. Since $\mathcal{D}_t$ depends on a triple correlation, it is harder to converge in experiments, and thus, its distributions appear less smooth, especially for the lower Reynolds number cases. The uncertainty in $\mathcal{P}_k$ is about 10\% (see Appendix~\ref{app:prod}).

 Previous studies have calculated these terms from experimental results for a step case \citep[e.g.][]{kasagi1995three, jovic1996experimental, piirto2003measuring}. While there are no measurements of axisymmetric expansions that directly compare to the range of parameters used in this study, the trends appear to be in good agreement with our measurements. The overall behavior of the TKE budget terms presented in Fig.~\ref{fig:turbsw_all} appears similar in all cases. However, for a given Reynolds number, both wedge cases consistently show higher production, with the peak shifted toward the free-stream. This trend indicates that turbulence is generated over a broader region in wedge cases, whereas in step cases, $\mathcal{P}_k$ remains concentrated near the step height due to the more localized shear layer development. These differences are closely linked to the fundamental shear layer dynamics that develop downstream of the expansion. The shear layer in wedge cases grows more rapidly and extends further into the free-stream compared to step cases as seen in Fig.~\ref{fig:shearlayer}(c). This is attributed to the gradual expansion allowing an attached return flow, leading to an increased interaction region between the high momentum free-stream and low-momentum return flow.  In contrast, the sharp separation in step cases confines the shear layer development to a smaller region, due to weaker interaction between the recirculating and incoming flow. The impingement angle and the momentum of the return flow is governed by the geometry of expansion (more in section~\ref{sec:conclusion}). Consequently,  $\mathcal{P}_k$, which depends on the Reynolds shear stresses and the mean velocity gradients (equation~\ref{eq:P}), is stronger and more distributed in wedge cases, whereas it remains concentrated near the corner in step cases.

 For $\mathcal{P}_k$, the lower Reynolds number cases ($Re_h$=25000) show higher magnitude, with W-25 reaching the highest peak. This suggests that at lower $Re_h$, the increased interaction in wedge cases generates larger fluctuations in the out-of-plane velocity, leading to higher out-of-plane Reynolds stresses seen in Fig.~\ref{fig:reynoldstress}(c). At $z^*$=4, W-35 exhibits a distinct peak near the step corner, likely due to enhanced turbulent mixing at the core of the primary recirculation region. A similar but attenuated pattern is observed in the other cases. Close to the reattachment zone, W-35 maintains the highest production, with a peak near $r^*$=1.9. Peaks for wedge cases are consistently located at $r^*>1.67$, whereas step cases peak at $r^*<1.67$, suggesting that the gradual expansion in wedge cases allows for a broader distribution of turbulence near the wall, as well as an elevated centerline for the shear layer. Additionally, the core of the large recirculation zone induces negative $\mathcal{P}_k$ close to the wall at $z^*$=4 in all cases. For $\mathcal{C}$, the $Re_h$=25000 cases exhibit higher magnitudes than the $Re_h$=35000 cases, with a pronounced negative peak close to $r^*$=1.5, followed by a smaller positive peak near the step height (Fig.~\ref{fig:turbsw_all}(b)).
 
 As the flow moves downstream, the negative peak in $\mathcal{C}$ shifts toward the centerline while decreasing in magnitude. This negative peak aligns with the outer extent of the shear layer, and the largest shear layer growth in W-25 shifts this peak further towards the centerline. Conversely, S-35, with the smallest shear layer growth, exhibits peaks closer to the wall at $z^*$=8. The progression of $\mathcal{C}$ further supports that gradual expansion enhances turbulence transport. For $\mathcal{D}$ shown in Fig.~\ref{fig:turbsw_all}(c), alternating negative and positive peaks appear at the edges of the shear layer, with the peaks separating and increasing in magnitude downstream. Large fluctuations observed in the recirculation region, especially in step cases, could be attributed to the presence of triple correlations in $\mathcal{D}_t$. These fluctuations suggest that high fluctuating velocities are not well converged in this region, except in the S-35 case, where turbulence transport appears more stable. The dissipation term, $\epsilon$, calculated as a residual, closely mirrors the trends observed in $\mathcal{P}_k$. The peak dissipation occurs at $r^* \approx 1.6$ for all cases, while it broadens and gradually decreases in magnitude downstream. The highest dissipation occurs at $z^*$=1 for W-25, at $z^*$=4 for W-35, and at $z^*$=8 for S-35, suggesting a strong dependency on expansion geometry and Reynolds number. 
 

 To further elucidate the spatial structure of turbulence generation near the expansion region, Fig.~\ref{fig:turbsw_all}(e) presents magnified maps of the TKE production term in the region immediately downstream of the step ($0.1 < z^* < 1.5$).  In all cases, production is highly localized along the upper edge of the recirculation region, coinciding with the developing shear layer. Notably, both wedge cases show broader and more intense production zones that extend significantly above the step height, reflecting the elevated shear and increased turbulence generation resulting from enhanced interaction between the return flow and the incoming free-stream. In contrast, the step cases exhibit more confined and lower-magnitude production regions, concentrated closer to the wall. The effect of the Reynolds number is also evident, with the $Re_h$=25000 cases showing larger $\mathcal{P}_k$ for both geometries. 

 These findings underscore the crucial role of shear layer dynamics in determining the observed differences in turbulence production between wedge and step geometries very close to the point of separation. For a given Reynolds number, the broader and more dispersed shear layer in wedge cases enhances $\mathcal{P}_k$, $\mathcal{D}$, and $\mathcal{C}$, resulting in higher and more sustained turbulence levels downstream. Conversely, in the step cases, the sharp separation limits the extent of turbulence generation, confining it closer to the step height and reducing its overall impact on the flow field. The results in Figs.~\ref{fig:TKE} and~\ref{fig:turbsw_all} agree well with previous observations, e.g., \cite{gibson1910flow} and \cite{idelchik1986handbook}, who showed an increase in energy losses in pipes with a 45$^\circ$ sloped expansion compared to a 90$^\circ$ abrupt expansion. While these results are widely used, the fundamental mechanism governing this behavior is not well understood.

\subsection{Reynolds stress anisotropy}
\label{ssec:anisotropy}

We employ the Reynolds stress anisotropy analysis to elucidate the physical mechanisms that govern the dynamics of flow separation and the distribution of turbulence intensities. \citet{choi2001return}  emphasized the importance of anisotropy in turbulence modeling, highlighting its impact on the energy cascade process. Experimental investigations, such as those by \citet{dey2020reynolds}, have provided detailed measurements of Reynolds stress anisotropy to identify different regions of the flow occurring near walls with significant undulations, offering valuable data for validating computational models. \citet{ fang2021direct} also used analysis of the Reynolds stress anisotropy to highlight the three-dimensionality of the flow on a forward-facing step, while \citet{jang2011effects} used it for axisymmetric  contraction in a pipe.  

\begin{figure}
    \centering
    \includegraphics[width=0.95\linewidth]{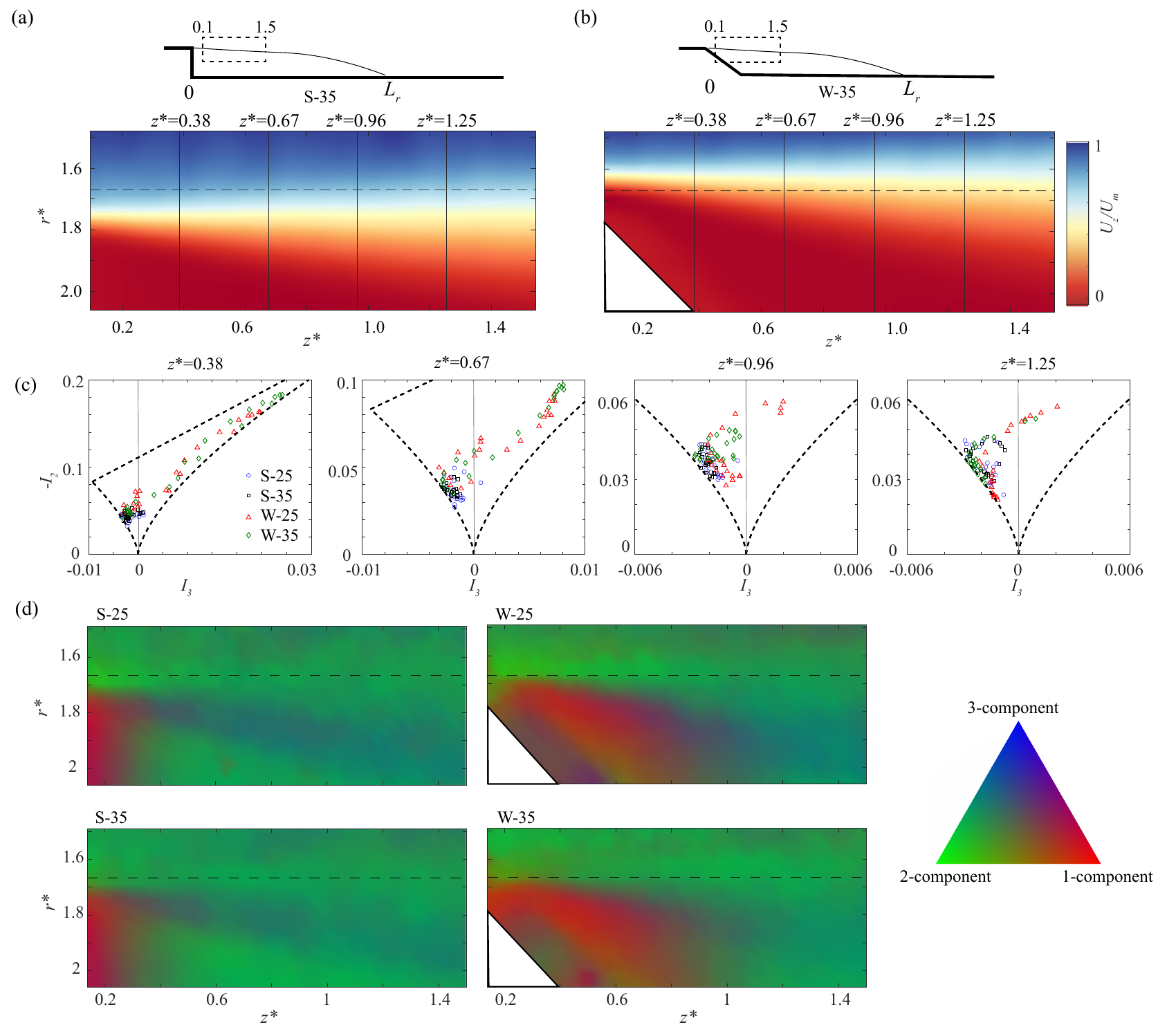}
    \caption{Flow fields near the step corner for (a) S-35 and (b) W-35 cases. A schematic of the region of interest is given above and the back vertical lines show the different $z^*$ locations. The horizontal dashed lines shows the location of the step edge. (c) Anisotropic invariant maps at various $z^*$ in the region shown in (a) and (b). The limits of the AIM are shown in dashed lines and the $I_3$=0 line in grey. (d)  Barycentric map of the AIM plotted above for all four cases in the region shown in (a) and (b). The triangle on the right shows the color scale as described by \cite{emory2014visualizing}. The dashed horizontal line represent the step height.}
    \label{fig:aniso}
\end{figure}

The degree of anisotropy in turbulence is evaluated using the Reynolds normal stresses $\sigma_z=\langle u_z'u_z' \rangle$, $\sigma_r=\langle u_r'u_r' \rangle$ and $\sigma_\theta=\langle u_\theta' u_\theta' \rangle$, where $\sigma_z=\sigma_r=\sigma_\theta$ represent isotropic turbulence. Using the Reynolds stress, we calculate the normalized anisotropy tensor $b_{ij}$ as 
\begin{equation}
    b_{ij}=\frac{\tau_{ij}}{2k}-\frac{\delta_{ij}}{3},
\end{equation}
where $\delta_{ij}$ is the kronecker delta. The Reynolds stress anisotropy tensor is symmetric and traceless, and the three principle invariants $I$ are related to $b_{ij}$ as, $I_1 =b_{kk}$, $I_2=-1/2(b_{ij}b_{ji})$ and $I_3=det(b_{ij})$. For incompressible flow, $I_1=0$, and $I_2=I_3=0$ for isotropic turbulence. For the flow to be axisymmetric, $I_3=\pm 2(-I_2/3)^{3/2}$. Using $-I_2$ and $I_3$, it is possible to plot the anisotropic invariant map (AIM), also known as Lumley triangle,  with the axisymmetric limits of $I_3$ forming the two curvilinear lines and topside linear line at the limits of $I_2$ and $I_3$ \citep{choi2001return,simonsen2005turbulent, dey2020reynolds}. The two-component axisymmetric limits are $-I_2$=1/12 and $I_3$=-1/108 while that of one-component is $-I_2$=1/3 and $I_3$=2/27. The bottom point of the AIM corresponds to isotropic turbulence with a spherical stress ellipsoid, while the left curved side is the axisymmetric contraction limit, where $\sigma_1=\sigma_2>\sigma_3$. The corresponding stress ellipsoid is an oblate spheroid, and at the two-component axisymmetric limit, it becomes a circular disk with $\sigma_1=\sigma_2$ and $\sigma_3=0$. Similarly, the right curved side is the axisymmetric expansion limit with $\sigma_1>\sigma_2=\sigma_3$, forming a prolate spheroid. Points which are located to the left of the $I_3=0$ line represent a shape of Reynolds stress ellipsoid which is an oblate spheroid i.e. in compression and those to the right represent a prolate spheroid i.e., in tension \citep{Pope_2000}. The linear top side is the two-component limit line corresponding to an elliptical disk. The axisymmetric contraction and expansion are symmetric about the plane-strain limit, where $I_3=0$. Finally, the right vertex corresponds to the one-component limit, with $\sigma_1>0$ and $\sigma_2=\sigma_3=0$. 

Fig.~\ref{fig:aniso} presents the results of the Reynolds stress anisotropy analysis performed close to the step ($z^*<1.25$) comparing all four cases. Figs.~\ref{fig:aniso}(a) and (b) show a magnified portion of this region, $0.1<z^*<1.5$, for S-35 and W-35, respectively. The step height is marked on the figures by a horizontal dashed line at $r^*$=1.67. Fig.~\ref{fig:aniso}(c) plots the AIM, using $-I_2$ and $I_3$ for sets of points taken along vertical lines located at four downstream locations marked on Figs.~\ref{fig:aniso}(a) and (b), comparing all cases. Significant differences in Reynolds stress anisotropy are mostly observed at the region close to the separation. The magnified view indicates that, in the wedge case, the high gradient region of the shear layer extends above the step height and into the free-stream, whereas for the step case, it remains below the step. This difference in dynamics of the shear layer corresponds with the results presented above, where the flow in the recirculation region of the wedge case is more organized and reaches the corner with more energy, thus pushing the flow into the core of the pipe. The results in Fig.~\ref{fig:aniso}(c) show that for the step cases, every point lies either very close to or to the left of $I_3$ = 0. In the wedge cases, locations that are far below the step height are similarly close to the left axisymmetric contraction limit as in the step case. However, closer to the core of the shear layer, their location on the AIM begins to shift, crossing the $I_3=0$ line, and the Reynolds stress ellipsoid becomes increasingly under tension. This disparity between the two geometries diminishes as the flow progresses downstream and the bulk of the shear layer moves downwards below the step height. The distinct change in the shape of the Reynolds stress ellipsoid, as seen in Fig.~\ref{fig:aniso}(c), unequivocally links the location of the shear layer in the wedge cases with increased interaction between it and the free-stream, leading to higher TKE production.

An alternative method to visualize and analyze the interaction between the return flow and the free stream is the barycentric map introduced by \citet{emory2014visualizing}. This approach provides a robust framework for visualizing turbulence anisotropy by mapping the Reynolds stress anisotropy tensor as a convex combination of three fundamental limiting states: one-component, two-component, and isotropic turbulence, represented as vertices of an equilateral triangle. The barycentric maps indicate that the incoming flow is almost isotropic throughout the bulk of the flow and becomes a two-component flow close to the walls. Additionally, downstream of the expansion, the region of isotropic flow expands along the centerline. Fig.~\ref{fig:aniso}(d) presents a comparison of the barycentric maps for all cases in the region shown in Figs.~\ref{fig:aniso}(a) and (b), providing a qualitative measure of the proximity of anisotropy tensor to these limiting behaviors using RGB color combination with red, green, and blue representing one, two and three component limits, respectively.  Most of the flow shows that there are two dominant Reynolds stress terms, which is to be expected due to the axisymmetric nature of the flow. A faint blue three-components streak can be seen in the lower parts of the shear layer where the vortical structures interact with the relatively slow flow in the recirculation region, leading to low momentum mixing and redistribution of the energy between the components. The barycentric map reveals additional major differences between the wedge and the step cases. As the return flow interacts with the free-stream, it decelerates and experiences compressive stresses. The affected region becomes dominated by one component (red regions in Fig.~\ref{fig:aniso}(d)). Comparing these barycentric maps with Fig.~\ref{fig:reynoldstress} identifies this component to be $\langle u_\theta' u_\theta' \rangle$ which increases significantly while the other stress components do not. These compression forces also indicate that the interaction between the return flow and the free-stream is significantly stronger in the wedge cases. Here, the compression leads to a form of `buckling' of the turbulence in the flow direction, which manifests as increased out-of-plane fluctuations. The barycentric maps show that the compressed region in the wedge cases is far larger than in the step cases and extends further into the recirculation region. For completeness, we have calculated the barycentric map upstream of the expansion, and the entire separated flow for the W-25 case is provided in Appendix~\ref{app:bary}. This indicates that, as one might expect, the turbulence in the core of the pipe remains predominantly isotropic. 

The Reynolds stress anisotropy further elucidates the mechanisms through which expansion geometry influences turbulence structure and energy transfer. In the wedge cases, the presence of a stronger and more organized return flow generates enhanced shear and drives elevated turbulence production near the expansion region. This is clearly evident in Fig.~\ref{fig:turbsw_all}(e), which shows that the production term peaks extend further downstream and reaches significantly higher magnitudes in the wedge cases, particularly W-25, compared to their step counterparts. These elevated production zones coincide with the regions of increased Reynolds shear stress, as shown in Fig.~\ref{fig:reynoldstress}, and are driven by the enhanced interaction between the low-momentum recirculating flow and the incoming free-stream. Additionally, the out-of-plane Reynolds stress component, $\langle u_\theta' u_\theta' \rangle$, remains higher in wedge cases (Fig.~\ref{fig:reynoldstress}(c)), supporting the interpretation that enhanced three-dimensional turbulence contributes to stronger energy redistribution in the early stages of flow development near the expansion corner. The corresponding anisotropy evolution, presented in Fig.~\ref{fig:aniso}, shows a clear departure from the axisymmetric contraction limit in wedge geometries, with trajectories on the anisotropic invariant map shifting toward more isotropic or even uniaxial configurations shortly after the expansion. This behavior indicates increased turbulence dimensionality and vortex stretching, driven by stronger and more spatially distributed production. In contrast, the step cases remain confined to lower production levels and more axisymmetric contraction states. Together, these findings demonstrate that the wedge cases not only amplifies the magnitude of turbulent energy production near the expansion but also alters the turbulence structure in a manner that promotes sustained mixing and the formation of larger coherent structures downstream.

\section{Discussions and Conclusions}
\label{sec:conclusion}

\begin{figure}
    \centering
    \includegraphics[width=\linewidth]{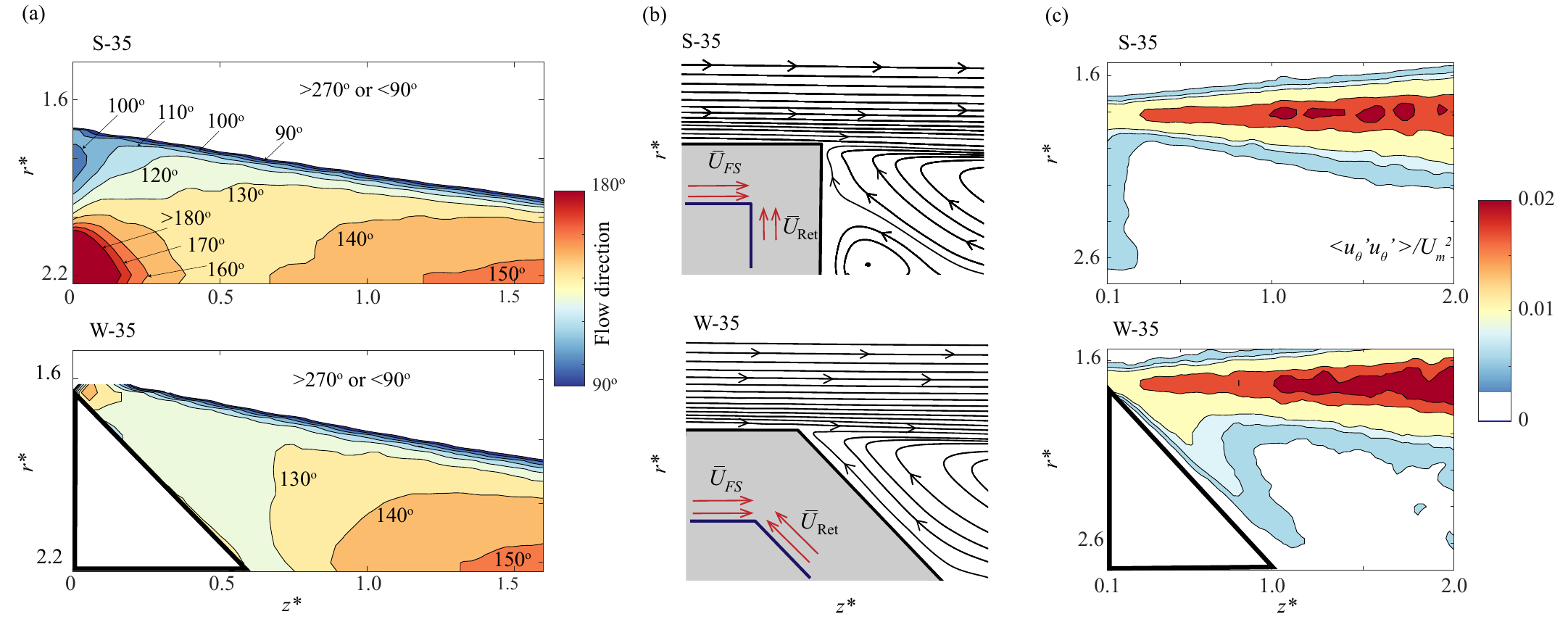}
    \caption{(a) Angle between $\langle U_z \rangle$ and $\langle U_r \rangle$ near the expansion for S-35 and W-35 cases, with angles $<90^\circ$ and $>270^\circ$ set to white for clarity (0$^\circ$: free-stream, 180$^\circ$: reverse flow).  (b) Sketch of recirculation effects on incoming flow for S-35 and W-35 cases with streamlines from mean flow data. (c) Magnified contour plot of normalized $\langle u_\theta' u_\theta' \rangle $ near the step corner for S-35 and W-35 cases, with magnitudes $<0.002$ set to white for easier interpretation.}
    \label{fig:sketch}
\end{figure}

The results presented above provide a comparative analysis of sharp and gradual expansions, revealing that the slope of the expansion surface fundamentally alters the organization of the return flow, which in turn dictates the structure and evolution of downstream turbulence. We have found that the wedge geometry facilitates a more organized return flow, which is aligned with the sloped surface, thereby preserving its momentum as it approaches the shear layer. Based on our findings, we have outlined the governing mechanism responsible for the distinct flow behaviors observed between the sharp and gradual expansions in Fig.~\ref{fig:sketch}. Fig.~\ref{fig:sketch}(a) compares the regions near the expansion, illustrating the direction of the flow vector in relation to the free-stream. This figure shows that in the wedge case, the flow maintains a uniform direction upstream as it moves along the slope until it abruptly interacts with the free-stream very close to the expansion corner. The flow turns sharply, resulting in very high local acceleration and shear, which leads to elevated production of turbulent kinetic energy near the expansion corner as seen in Fig.~\ref{fig:turbsw_all}(e). Conversely, Fig.~\ref{fig:sketch}(a) shows that in the step geometry, the flow approaches the free-stream with a far smaller relative angle, which leads to weaker interaction with the free-stream.  Furthermore, the formation of a secondary vortex ahead of the wall causes the return flow to separate from the surface and recirculate, further reducing its momentum.

Fig.~\ref{fig:sketch}(b) presents a comparison of the streamlines close to the area expansion, highlighting the direction of the return flow as it approaches the incoming free-stream. The figure schematically shows the velocity vectors of the incoming free-stream, $\overline{U}_{\it FS}$, and the return flow, $\overline{U}_{\it Ret}$, near the separation point. In wedge geometries, $\overline{U}_{\it Ret}$ retains a relatively high $-z$ component that opposes the free-stream direction, thus significantly increasing the interaction forces between the return flow and the free-stream. While our study does not examine the effect of different angles, it aligns with, and intuitively explains previous observations which showed that in cases of fully separated flow across area expansions, the energy losses peak at a slope angle of $\sim 30^\circ$ (see Fig.~\ref{fig:bfs_schematic}(b)) and reduces as the angle of the slope increases to $ 90^\circ$. In the latter, the $z$ component of $\overline{U}_{\it Ret}$ goes to zero. The figure shows that in the step case, the geometry directs the return flow upwards, suppressing horizontal momentum transfer, and the change in direction of the return flow is not as sharp as in the wedge case. The increased interaction facilitated by the angle of the return flow leads to stronger shear in the wedge cases, which explains the increased turbulence production seen in Fig.~\ref{fig:turbsw_all}. Additionally, Fig.~\ref{fig:sketch}(c) shows the formation of a significant out-of-plane Reynolds stress component, $\langle u_\theta' u_\theta' \rangle$,  in the wedge case. This occurs due to an increase in compressive stresses resulting from the impingement of the stronger reverse flow on the free-stream in the wedge case, which induces out-of-plane flow. The mechanism described here also explains the findings presented in Fig.~\ref{fig:aniso}(a), where the increased compressive forces push the shear layer into the free-stream, leading to distinct changes in the morphology of the Anisotropic invariant maps. The increase in the out-of-plane component explains the transition away from the axisymmetric contraction limit toward a more three-component or uniaxial turbulence state, as shown in Fig.~\ref{fig:aniso}(c). 

While it has been known that sloped expansion leads to increased turbulence and energy losses, our findings elucidate the fundamental mechanism and elaborate on the flow features that are affected by the expansion slope. The expansion geometry governs the turbulence by setting the strength and alignment of the return flow. The gradual expansion promotes strong shear-layer interactions, elevated TKE production, and increased out-of-plane flow in the immediate vicinity of the area expansion, all of which contribute to enhanced turbulence. In contrast, the abrupt expansion leads to a reduced interaction of the return flow and suppresses turbulence production. By comparing turbulent flow through sloped ($45^\circ$) and abrupt ($90^\circ$) axisymmetric expansions, this study elucidates how expansion geometry governs return flow dynamics and shear layer development. The results offer a mechanistic explanation for the increased losses observed in sloped expansions and provide high-fidelity data for turbulence model validation. While only two expansion angles were examined, the consistency and clarity of the observed trends suggest that the proposed mechanism is robust and likely extends to a broader range of geometries and flow conditions, warranting further investigation. 

\section*{Acknowledgments}

J.T.Jose is partially supported by a Technion Post-Doctoral fellowship.  The authors also wish to express their gratitude to Prof. Joseph Katz at Johns Hopkins University for his valuable suggestions and insightful contributions during the discussions of our findings.
This work was supported by the donors of ACS Petroleum Research Fund under New Directions Grant \#65901-ND9

\appendix
\section{Uncertainty Estimation in Stereo PIV}
\label{app:Uncertainty}

The uncertainty quantification method is based on the statistical evaluation of the correlation peak shape in the cross-correlation process. In an ideal, noise-free measurement, particle image pairs in interrogation windows would align perfectly, resulting in a symmetric correlation peak. However, in real experimental conditions, noise, optical distortions, and calibration errors introduce asymmetries in the correlation peak, from which the uncertainty of the displacement vector is derived \citep{wieneke_piv_2017}. For stereo PIV, the velocity components were reconstructed from the 2D displacement fields obtained from each camera. The uncertainty in the measured displacement vectors from each view was propagated through the stereo reconstruction equations to obtain the uncertainties in the three velocity components ($U_z, U_r, U_{\theta}$). The propagation follows the standard approach for uncertainty in a derived quantity $y$, given as a function of $N$ measured variables $x_i$ \citep{sciacchitano2016piv}:

\begin{equation}
\sigma_y^2 = \sum_{i=1}^{N} \left( \frac{\partial y}{\partial x_i} \sigma_{x_i} \right)^2 + 2 \sum_{i=1}^{N} \sum_{j=i+1}^{N} \left( \frac{\partial y}{\partial x_i} \frac{\partial y}{\partial x_j} \rho_{x_i x_j} \sigma_{x_i} \sigma_{x_j} \right),
\end{equation}

where $\sigma_{x_i}$ is the uncertainty in each measured variable, and $\rho_{x_i x_j}$ is the correlation coefficient between two variables. In stereo PIV, the uncertainties in the apparent displacements from each camera ($V_{x1}, V_{y1}, V_{x2}, V_{y2}$) contribute to the final velocity uncertainties, considering their inter-dependencies.

For a stereo setup with two cameras viewing at angles $\alpha_1$ and $\alpha_2$, the uncertainties in the reconstructed velocity components propagate as follows \citep{wieneke_piv_2017}:

\begin{equation}
\sigma_U = \frac{\sigma_{V_{x1}, V_{x2}}}{\sqrt{2} |\cos\alpha|},
\end{equation}

\begin{equation}
\sigma_W = \frac{\sigma_{V_{x1}, V_{x2}}}{\sqrt{2} |\sin\alpha|},
\end{equation}

where $\sigma_{V_{x1}, V_{x2}}$ represents the uncertainty in the apparent in-plane velocity components from each camera. Given the stereo camera arrangement used in this study, strong correlations were observed between the $U_z$ and $U_{\theta}$ components, which were accounted for in the uncertainty estimation. As per this approach, the estimated maximum uncertainty in the mean velocity components were 0.4\% of $U_m$ for $U_z$, 0.25\% for $U_r$, and 0.3\% for $U_{\theta}$ for a data set with $N$=3000 images. The above analysis was performed using built-in tools available in LaVision DaVis$^{\text{TM}}$ 11 PIV software.

Additionally, the uncertainty in derived quantities, such as Reynolds stresses, is estimated using standard error propagation techniques introduced in \cite{sciacchitano2016piv}. Given velocity components $u$ and $v$ with known uncertainties $\sigma_u$ and $\sigma_v$, the uncertainty in the Reynolds stress $\overline{u' v'}$ is given by:

\begin{equation}
\sigma_{\overline{u' v'}} = \sqrt{ (\sigma_u v)^2 + (\sigma_v u)^2 + 2 \cdot \text{Cov}(u, v)}
\end{equation}

where $\text{Cov}(u, v)$ represents the covariance between velocity components (Reynolds stress). In stereo PIV, uncertainty propagation considers correlations between components to ensure accurate quantification of measurement uncertainties. The uncertainty estimate for mean TKE is 0.35\% of $U_m^2$, while that of Reynolds stress components are less than 0.15\% of $U_m^2$ for all the terms.

\section{Uncertainty Estimation in Turbulence Production}
\label{app:prod}

The uncertainty in the turbulence production term $\mathcal{P}_k$ can be estimated using the propagation of uncertainty for products of independently measured quantities. The production term is calculated as $\mathcal{P}_k = -\langle u'_i u'_j \rangle \, \partial \langle U_i \rangle / \partial x_j$, and its uncertainty is estimated as:

\begin{equation}
    \frac{\delta \mathcal{P}_k}{\mathcal{P}_k} = \sqrt{\left( \frac{\delta \langle u'_i u'_j \rangle}{\langle u'_i u'_j \rangle} \right)^2 + \left( \frac{\delta (\partial \langle U_i \rangle / \partial x_j)}{\partial \langle U_i \rangle / \partial x_j} \right)^2}
\end{equation}

The spatial resolution of the stereo PIV data is $0.4953$ mm, and the uncertainty in the mean velocity components is less than 0.4\% of the inlet bulk velocity $U_m = 5 \, \text{m/s}$. This results in gradient uncertainties of approximately $28.6\text{s}^{-1}$ in $\partial \langle U_z \rangle / \partial z$, $17.9\text{s}^{-1}$ in $\partial \langle U_r \rangle / \partial r$, and $40.5 \, \text{s}^{-1}$ in the combined gradient $\partial \langle U_z \rangle / \partial r + \partial \langle U_r \rangle / \partial z$. Reynolds stress uncertainties are taken as $0.0375\text{m}^2/\text{s}^2$, corresponding to 0.15\% of $U_m^2$. Using representative values of the Reynolds stress and mean velocity gradients in the shear layer, the resulting relative uncertainties in individual production terms ranges from 7.7\% to 15.7\%, with the total uncertainty in $\mathcal{P}_k$ estimated to be approximately 10.7\%.

\section{Upstream Turbulence Anisotropy and Barycentric Mapping}
\label{app:bary}

\begin{figure}[h]
    \centering
    \includegraphics[width=0.95\linewidth]{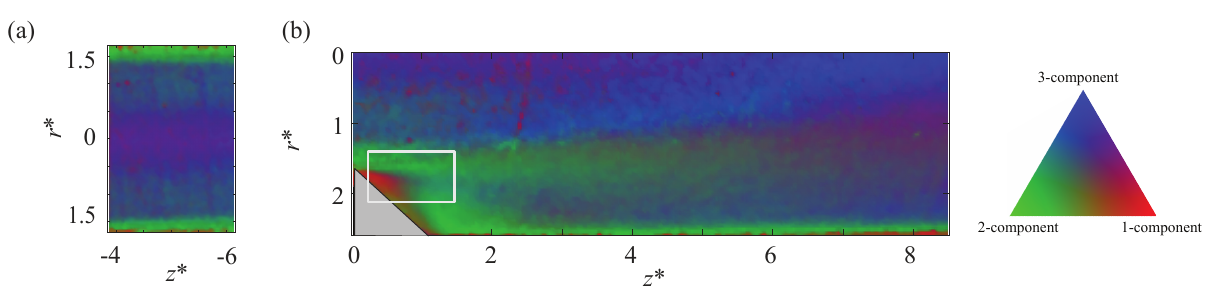}
    \caption{Barycentric anisotropy maps for the W-25 case (a) upstream of the expansion, showing isotropic core and near-wall two-component turbulence, and (b) downstream expansion region, with the white box marking the subregion used in Fig.~\ref{fig:aniso}}
    \label{fig:appendix_bary}
\end{figure}

The anisotropy of turbulence upstream of the pipe expansion is evaluated using barycentric maps to verify the validity of the fully developed turbulent inlet condition. As shown in Fig.~\ref{fig:appendix_bary}(a), which corresponds to a region well upstream of the expansion, turbulence in the pipe core remains predominantly isotropic, consistent with expectations for fully developed turbulent pipe flow. A transition to two-component turbulence appears only near the wall, in agreement with canonical DNS results for internal turbulent flows \citep[e.g.,][]{emory2014visualizing}. For comparison, Fig.\ref{fig:appendix_bary}(b) presents the barycentric map downstream of the expansion (corresponding to Fig.\ref{fig:aniso}(d)), with the region shown in the main manuscript outlined by a white box. In this region, complex anisotropy patterns emerge due to shear layer development and flow separation. However, the upstream map clearly demonstrates that the incoming flow remains undistorted by the upstream geometry, and the inflow condition is three-dimensionally isotropic in the bulk. These findings confirm that the downstream anisotropy reflects physical mechanisms rather than artifacts of the test section design.

\bibliography{references}

\begin{thebibliography}{70}%
\makeatletter
\providecommand \@ifxundefined [1]{%
 \@ifx{#1\undefined}
}%
\providecommand \@ifnum [1]{%
 \ifnum #1\expandafter \@firstoftwo
 \else \expandafter \@secondoftwo
 \fi
}%
\providecommand \@ifx [1]{%
 \ifx #1\expandafter \@firstoftwo
 \else \expandafter \@secondoftwo
 \fi
}%
\providecommand \natexlab [1]{#1}%
\providecommand \enquote  [1]{``#1''}%
\providecommand \bibnamefont  [1]{#1}%
\providecommand \bibfnamefont [1]{#1}%
\providecommand \citenamefont [1]{#1}%
\providecommand \href@noop [0]{\@secondoftwo}%
\providecommand \href [0]{\begingroup \@sanitize@url \@href}%
\providecommand \@href[1]{\@@startlink{#1}\@@href}%
\providecommand \@@href[1]{\endgroup#1\@@endlink}%
\providecommand \@sanitize@url [0]{\catcode `\\12\catcode `\$12\catcode `\&12\catcode `\#12\catcode `\^12\catcode `\_12\catcode `\%12\relax}%
\providecommand \@@startlink[1]{}%
\providecommand \@@endlink[0]{}%
\providecommand \url  [0]{\begingroup\@sanitize@url \@url }%
\providecommand \@url [1]{\endgroup\@href {#1}{\urlprefix }}%
\providecommand \urlprefix  [0]{URL }%
\providecommand \Eprint [0]{\href }%
\providecommand \doibase [0]{https://doi.org/}%
\providecommand \selectlanguage [0]{\@gobble}%
\providecommand \bibinfo  [0]{\@secondoftwo}%
\providecommand \bibfield  [0]{\@secondoftwo}%
\providecommand \translation [1]{[#1]}%
\providecommand \BibitemOpen [0]{}%
\providecommand \bibitemStop [0]{}%
\providecommand \bibitemNoStop [0]{.\EOS\space}%
\providecommand \EOS [0]{\spacefactor3000\relax}%
\providecommand \BibitemShut  [1]{\csname bibitem#1\endcsname}%
\let\auto@bib@innerbib\@empty
\bibitem [{\citenamefont {Eaton}(1980)}]{eaton1980turbulent}%
  \BibitemOpen
  \bibfield  {author} {\bibinfo {author} {\bibfnamefont {J.~K.}\ \bibnamefont {Eaton}},\ }\href@noop {} {\emph {\bibinfo {title} {Turbulent flow reattachment: an experimental study of the flow and structure behind a backward-facing step}}}\ (\bibinfo  {publisher} {Stanford University},\ \bibinfo {year} {1980})\BibitemShut {NoStop}%
\bibitem [{\citenamefont {{\"O}t{\"u}gen}(1991)}]{otugen1991expansion}%
  \BibitemOpen
  \bibfield  {author} {\bibinfo {author} {\bibfnamefont {M.}~\bibnamefont {{\"O}t{\"u}gen}},\ }\bibfield  {title} {\bibinfo {title} {Expansion ratio effects on the separated shear layer and reattachment downstream of a backward-facing step},\ }\href@noop {} {\bibfield  {journal} {\bibinfo  {journal} {Experiments in fluids}\ }\textbf {\bibinfo {volume} {10}},\ \bibinfo {pages} {273} (\bibinfo {year} {1991})}\BibitemShut {NoStop}%
\bibitem [{\citenamefont {Le}\ \emph {et~al.}(1997)\citenamefont {Le}, \citenamefont {Moin},\ and\ \citenamefont {Kim}}]{le1997direct}%
  \BibitemOpen
  \bibfield  {author} {\bibinfo {author} {\bibfnamefont {H.}~\bibnamefont {Le}}, \bibinfo {author} {\bibfnamefont {P.}~\bibnamefont {Moin}},\ and\ \bibinfo {author} {\bibfnamefont {J.}~\bibnamefont {Kim}},\ }\bibfield  {title} {\bibinfo {title} {Direct numerical simulation of turbulent flow over a backward-facing step},\ }\href@noop {} {\bibfield  {journal} {\bibinfo  {journal} {Journal of fluid mechanics}\ }\textbf {\bibinfo {volume} {330}},\ \bibinfo {pages} {349} (\bibinfo {year} {1997})}\BibitemShut {NoStop}%
\bibitem [{\citenamefont {Nadge}\ and\ \citenamefont {Govardhan}(2014)}]{nadge2014high}%
  \BibitemOpen
  \bibfield  {author} {\bibinfo {author} {\bibfnamefont {P.~M.}\ \bibnamefont {Nadge}}\ and\ \bibinfo {author} {\bibfnamefont {R.}~\bibnamefont {Govardhan}},\ }\bibfield  {title} {\bibinfo {title} {High reynolds number flow over a backward-facing step: structure of the mean separation bubble},\ }\href@noop {} {\bibfield  {journal} {\bibinfo  {journal} {Experiments in fluids}\ }\textbf {\bibinfo {volume} {55}},\ \bibinfo {pages} {1} (\bibinfo {year} {2014})}\BibitemShut {NoStop}%
\bibitem [{\citenamefont {Eaton}\ and\ \citenamefont {Johnston}(1982)}]{eaton1982low}%
  \BibitemOpen
  \bibfield  {author} {\bibinfo {author} {\bibfnamefont {J.~K.}\ \bibnamefont {Eaton}}\ and\ \bibinfo {author} {\bibfnamefont {J.~P.}\ \bibnamefont {Johnston}},\ }\bibfield  {title} {\bibinfo {title} {Low frequency unsteadyness of a reattaching turbulent shear layer},\ }in\ \href@noop {} {\emph {\bibinfo {booktitle} {Turbulent Shear Flows 3: Selected Papers from the Third International Symposium on Turbulent Shear Flows, The University of California, Davis, September 9--11, 1981}}}\ (\bibinfo {organization} {Springer},\ \bibinfo {year} {1982})\ pp.\ \bibinfo {pages} {162--170}\BibitemShut {NoStop}%
\bibitem [{\citenamefont {Scharnowski}(2014)}]{scharnowski2014investigation}%
  \BibitemOpen
  \bibfield  {author} {\bibinfo {author} {\bibfnamefont {S.}~\bibnamefont {Scharnowski}},\ }\emph {\bibinfo {title} {Investigation of turbulent shear flows with high resolution PIV methods}},\ \href@noop {} {Ph.D. thesis},\ \bibinfo  {school} {M{\"u}nchen, Univ. der Bundeswehr, Diss., 2013} (\bibinfo {year} {2014})\BibitemShut {NoStop}%
\bibitem [{\citenamefont {McKeon}\ \emph {et~al.}(2004)\citenamefont {McKeon}, \citenamefont {Swanson}, \citenamefont {Zagarola}, \citenamefont {Donnelly},\ and\ \citenamefont {Smits}}]{mckeon2004friction}%
  \BibitemOpen
  \bibfield  {author} {\bibinfo {author} {\bibfnamefont {B.}~\bibnamefont {McKeon}}, \bibinfo {author} {\bibfnamefont {C.}~\bibnamefont {Swanson}}, \bibinfo {author} {\bibfnamefont {M.}~\bibnamefont {Zagarola}}, \bibinfo {author} {\bibfnamefont {R.}~\bibnamefont {Donnelly}},\ and\ \bibinfo {author} {\bibfnamefont {A.~J.}\ \bibnamefont {Smits}},\ }\bibfield  {title} {\bibinfo {title} {Friction factors for smooth pipe flow},\ }\href@noop {} {\bibfield  {journal} {\bibinfo  {journal} {Journal of Fluid Mechanics}\ }\textbf {\bibinfo {volume} {511}},\ \bibinfo {pages} {41} (\bibinfo {year} {2004})}\BibitemShut {NoStop}%
\bibitem [{\citenamefont {Marusic}\ \emph {et~al.}(2010)\citenamefont {Marusic}, \citenamefont {McKeon}, \citenamefont {Monkewitz}, \citenamefont {Nagib}, \citenamefont {Smits},\ and\ \citenamefont {Sreenivasan}}]{marusic2010wall}%
  \BibitemOpen
  \bibfield  {author} {\bibinfo {author} {\bibfnamefont {I.}~\bibnamefont {Marusic}}, \bibinfo {author} {\bibfnamefont {B.~J.}\ \bibnamefont {McKeon}}, \bibinfo {author} {\bibfnamefont {P.~A.}\ \bibnamefont {Monkewitz}}, \bibinfo {author} {\bibfnamefont {H.~M.}\ \bibnamefont {Nagib}}, \bibinfo {author} {\bibfnamefont {A.~J.}\ \bibnamefont {Smits}},\ and\ \bibinfo {author} {\bibfnamefont {K.~R.}\ \bibnamefont {Sreenivasan}},\ }\bibfield  {title} {\bibinfo {title} {Wall-bounded turbulent flows at high reynolds numbers: recent advances and key issues},\ }\href@noop {} {\bibfield  {journal} {\bibinfo  {journal} {Physics of fluids}\ }\textbf {\bibinfo {volume} {22}} (\bibinfo {year} {2010})}\BibitemShut {NoStop}%
\bibitem [{\citenamefont {Gibson}(1910)}]{gibson1910flow}%
  \BibitemOpen
  \bibfield  {author} {\bibinfo {author} {\bibfnamefont {A.}~\bibnamefont {Gibson}},\ }\bibfield  {title} {\bibinfo {title} {On the flow of water through pipes and passages having converging or diverging boundaries},\ }\href@noop {} {\bibfield  {journal} {\bibinfo  {journal} {Proceedings of the Royal Society of London. Series A, Containing Papers of a Mathematical and Physical Character}\ }\textbf {\bibinfo {volume} {83}},\ \bibinfo {pages} {366} (\bibinfo {year} {1910})}\BibitemShut {NoStop}%
\bibitem [{\citenamefont {Idelchik}(1986)}]{idelchik1986handbook}%
  \BibitemOpen
  \bibfield  {author} {\bibinfo {author} {\bibfnamefont {I.~E.}\ \bibnamefont {Idelchik}},\ }\href@noop {} {\emph {\bibinfo {title} {Handbook of hydraulic resistance}}},\ \bibinfo {edition} {4th}\ ed.\ (\bibinfo  {publisher} {Begell House, New York},\ \bibinfo {year} {1986})\BibitemShut {NoStop}%
\bibitem [{\citenamefont {White}(1998)}]{White1998}%
  \BibitemOpen
  \bibfield  {author} {\bibinfo {author} {\bibfnamefont {F.~M.}\ \bibnamefont {White}},\ }\href@noop {} {\emph {\bibinfo {title} {Fluid Mechanics}}},\ \bibinfo {edition} {4th}\ ed.\ (\bibinfo  {publisher} {McGraw-Hill},\ \bibinfo {address} {New York, NY},\ \bibinfo {year} {1998})\BibitemShut {NoStop}%
\bibitem [{\citenamefont {Hammad}\ \emph {et~al.}(1999{\natexlab{a}})\citenamefont {Hammad}, \citenamefont {{\"O}t{\"u}gen},\ and\ \citenamefont {Arik}}]{hammad1999piv}%
  \BibitemOpen
  \bibfield  {author} {\bibinfo {author} {\bibfnamefont {K.~J.}\ \bibnamefont {Hammad}}, \bibinfo {author} {\bibfnamefont {M.~V.}\ \bibnamefont {{\"O}t{\"u}gen}},\ and\ \bibinfo {author} {\bibfnamefont {E.~B.}\ \bibnamefont {Arik}},\ }\bibfield  {title} {\bibinfo {title} {A piv study of the laminar axisymmetric sudden expansion flow},\ }\href@noop {} {\bibfield  {journal} {\bibinfo  {journal} {Experiments in fluids}\ }\textbf {\bibinfo {volume} {26}},\ \bibinfo {pages} {266} (\bibinfo {year} {1999}{\natexlab{a}})}\BibitemShut {NoStop}%
\bibitem [{\citenamefont {Amini}\ and\ \citenamefont {Hassan}(2012)}]{amini2012investigation}%
  \BibitemOpen
  \bibfield  {author} {\bibinfo {author} {\bibfnamefont {N.}~\bibnamefont {Amini}}\ and\ \bibinfo {author} {\bibfnamefont {Y.~A.}\ \bibnamefont {Hassan}},\ }\bibfield  {title} {\bibinfo {title} {An investigation of matched index of refraction technique and its application in optical measurements of fluid flow},\ }\href@noop {} {\bibfield  {journal} {\bibinfo  {journal} {Experiments in fluids}\ }\textbf {\bibinfo {volume} {53}},\ \bibinfo {pages} {2011} (\bibinfo {year} {2012})}\BibitemShut {NoStop}%
\bibitem [{\citenamefont {Mak}\ and\ \citenamefont {Balabani}(2007)}]{mak2007near}%
  \BibitemOpen
  \bibfield  {author} {\bibinfo {author} {\bibfnamefont {H.}~\bibnamefont {Mak}}\ and\ \bibinfo {author} {\bibfnamefont {S.}~\bibnamefont {Balabani}},\ }\bibfield  {title} {\bibinfo {title} {Near field characteristics of swirling flow past a sudden expansion},\ }\href@noop {} {\bibfield  {journal} {\bibinfo  {journal} {Chemical engineering science}\ }\textbf {\bibinfo {volume} {62}},\ \bibinfo {pages} {6726} (\bibinfo {year} {2007})}\BibitemShut {NoStop}%
\bibitem [{\citenamefont {Goharzadeh}\ and\ \citenamefont {Rodgers}(2009)}]{goharzadeh2009experimental}%
  \BibitemOpen
  \bibfield  {author} {\bibinfo {author} {\bibfnamefont {A.}~\bibnamefont {Goharzadeh}}\ and\ \bibinfo {author} {\bibfnamefont {P.}~\bibnamefont {Rodgers}},\ }\bibfield  {title} {\bibinfo {title} {Experimental measurement of laminar axisymmetric flow through confined annular geometries with sudden inward expansion},\ }\href@noop {} {\bibfield  {journal} {\bibinfo  {journal} {Journal of Fluids Engineering}\ }\textbf {\bibinfo {volume} {131}},\ \bibinfo {pages} {124501} (\bibinfo {year} {2009})}\BibitemShut {NoStop}%
\bibitem [{\citenamefont {Hammad}\ \emph {et~al.}(1999{\natexlab{b}})\citenamefont {Hammad}, \citenamefont {{\"O}t{\"u}gen}, \citenamefont {Vradis},\ and\ \citenamefont {Arik}}]{hammad1999laminar}%
  \BibitemOpen
  \bibfield  {author} {\bibinfo {author} {\bibfnamefont {K.~J.}\ \bibnamefont {Hammad}}, \bibinfo {author} {\bibfnamefont {M.~V.}\ \bibnamefont {{\"O}t{\"u}gen}}, \bibinfo {author} {\bibfnamefont {G.~C.}\ \bibnamefont {Vradis}},\ and\ \bibinfo {author} {\bibfnamefont {E.~B.}\ \bibnamefont {Arik}},\ }\bibfield  {title} {\bibinfo {title} {Laminar flow of a nonlinear viscoplastic fluid through an axisymmetric sudden expansion},\ }\href@noop {} {\bibfield  {journal} {\bibinfo  {journal} {Journal of Fluids Engineering}\ } (\bibinfo {year} {1999}{\natexlab{b}})}\BibitemShut {NoStop}%
\bibitem [{\citenamefont {Back}\ and\ \citenamefont {Roschke}(1972)}]{back1972shear}%
  \BibitemOpen
  \bibfield  {author} {\bibinfo {author} {\bibfnamefont {L.~H.}\ \bibnamefont {Back}}\ and\ \bibinfo {author} {\bibfnamefont {E.~J.}\ \bibnamefont {Roschke}},\ }\bibfield  {title} {\bibinfo {title} {Shear-layer flow regimes and wave instabilities and reattachment lengths downstream of an abrupt circular channel expansion},\ }\href@noop {} {\bibfield  {journal} {\bibinfo  {journal} {Journal of Applied Mechanics}\ }\textbf {\bibinfo {volume} {39}},\ \bibinfo {pages} {677} (\bibinfo {year} {1972})}\BibitemShut {NoStop}%
\bibitem [{\citenamefont {Devenport}\ and\ \citenamefont {Sutton}(1993)}]{devenport1993experimental}%
  \BibitemOpen
  \bibfield  {author} {\bibinfo {author} {\bibfnamefont {W.}~\bibnamefont {Devenport}}\ and\ \bibinfo {author} {\bibfnamefont {E.}~\bibnamefont {Sutton}},\ }\bibfield  {title} {\bibinfo {title} {An experimental study of two flows through an axisymmetric sudden expansion},\ }\href@noop {} {\bibfield  {journal} {\bibinfo  {journal} {Experiments in Fluids}\ }\textbf {\bibinfo {volume} {14}},\ \bibinfo {pages} {423} (\bibinfo {year} {1993})}\BibitemShut {NoStop}%
\bibitem [{\citenamefont {Morrison}\ \emph {et~al.}(1988)\citenamefont {Morrison}, \citenamefont {Tatterson},\ and\ \citenamefont {Long}}]{morrison1988three}%
  \BibitemOpen
  \bibfield  {author} {\bibinfo {author} {\bibfnamefont {G.}~\bibnamefont {Morrison}}, \bibinfo {author} {\bibfnamefont {G.~B.}\ \bibnamefont {Tatterson}},\ and\ \bibinfo {author} {\bibfnamefont {M.}~\bibnamefont {Long}},\ }\bibfield  {title} {\bibinfo {title} {Three-dimensional laser velocimeter investigation of turbulent, incompressible flow in an axisymmetric sudden expansion},\ }\href@noop {} {\bibfield  {journal} {\bibinfo  {journal} {Journal of Propulsion and Power}\ }\textbf {\bibinfo {volume} {4}},\ \bibinfo {pages} {533} (\bibinfo {year} {1988})}\BibitemShut {NoStop}%
\bibitem [{\citenamefont {Durrett}\ \emph {et~al.}(1988)\citenamefont {Durrett}, \citenamefont {Stevenson},\ and\ \citenamefont {Thompson}}]{durrett1988radial}%
  \BibitemOpen
  \bibfield  {author} {\bibinfo {author} {\bibfnamefont {R.}~\bibnamefont {Durrett}}, \bibinfo {author} {\bibfnamefont {W.}~\bibnamefont {Stevenson}},\ and\ \bibinfo {author} {\bibfnamefont {H.}~\bibnamefont {Thompson}},\ }\bibfield  {title} {\bibinfo {title} {Radial and axial turbulent flow measurements with an ldv in an axisymmetric sudden expansion air flow},\ }\href@noop {} {\bibfield  {journal} {\bibinfo  {journal} {Journal of Fluids Engineering}\ } (\bibinfo {year} {1988})}\BibitemShut {NoStop}%
\bibitem [{\citenamefont {Stieglmeier}\ \emph {et~al.}(1989)\citenamefont {Stieglmeier}, \citenamefont {Tropea}, \citenamefont {Weiser},\ and\ \citenamefont {Nitsche}}]{stieglmeier1989experimental}%
  \BibitemOpen
  \bibfield  {author} {\bibinfo {author} {\bibfnamefont {M.}~\bibnamefont {Stieglmeier}}, \bibinfo {author} {\bibfnamefont {C.}~\bibnamefont {Tropea}}, \bibinfo {author} {\bibfnamefont {N.}~\bibnamefont {Weiser}},\ and\ \bibinfo {author} {\bibfnamefont {W.}~\bibnamefont {Nitsche}},\ }\bibfield  {title} {\bibinfo {title} {Experimental investigation of the flow through axisymmetric expansions},\ }\href@noop {} {\bibfield  {journal} {\bibinfo  {journal} {Journal of Fluids Engineering}\ }\textbf {\bibinfo {volume} {111}},\ \bibinfo {pages} {464} (\bibinfo {year} {1989})}\BibitemShut {NoStop}%
\bibitem [{\citenamefont {Gould}\ \emph {et~al.}(1990)\citenamefont {Gould}, \citenamefont {Stevenson},\ and\ \citenamefont {Thompson}}]{gould1990investigation}%
  \BibitemOpen
  \bibfield  {author} {\bibinfo {author} {\bibfnamefont {R.~D.}\ \bibnamefont {Gould}}, \bibinfo {author} {\bibfnamefont {W.~H.}\ \bibnamefont {Stevenson}},\ and\ \bibinfo {author} {\bibfnamefont {H.~D.}\ \bibnamefont {Thompson}},\ }\bibfield  {title} {\bibinfo {title} {Investigation of turbulent transport in an axisymmetric sudden expansion},\ }\href@noop {} {\bibfield  {journal} {\bibinfo  {journal} {AIAA journal}\ }\textbf {\bibinfo {volume} {28}},\ \bibinfo {pages} {276} (\bibinfo {year} {1990})}\BibitemShut {NoStop}%
\bibitem [{\citenamefont {DeOtte~Jr}\ \emph {et~al.}(1991)\citenamefont {DeOtte~Jr}, \citenamefont {Morrison}, \citenamefont {Panak},\ and\ \citenamefont {Nail}}]{deotte19913}%
  \BibitemOpen
  \bibfield  {author} {\bibinfo {author} {\bibfnamefont {R.}~\bibnamefont {DeOtte~Jr}}, \bibinfo {author} {\bibfnamefont {G.}~\bibnamefont {Morrison}}, \bibinfo {author} {\bibfnamefont {D.}~\bibnamefont {Panak}},\ and\ \bibinfo {author} {\bibfnamefont {G.}~\bibnamefont {Nail}},\ }\bibfield  {title} {\bibinfo {title} {3-d laser doppler anemometry measurements of the axisymmetric flow field near an orifice plate},\ }\href@noop {} {\bibfield  {journal} {\bibinfo  {journal} {Flow Measurement and Instrumentation}\ }\textbf {\bibinfo {volume} {2}},\ \bibinfo {pages} {115} (\bibinfo {year} {1991})}\BibitemShut {NoStop}%
\bibitem [{\citenamefont {Furuichi}\ \emph {et~al.}(2003)\citenamefont {Furuichi}, \citenamefont {Takeda},\ and\ \citenamefont {Kumada}}]{furuichi2003spatial}%
  \BibitemOpen
  \bibfield  {author} {\bibinfo {author} {\bibfnamefont {N.}~\bibnamefont {Furuichi}}, \bibinfo {author} {\bibfnamefont {Y.}~\bibnamefont {Takeda}},\ and\ \bibinfo {author} {\bibfnamefont {M.}~\bibnamefont {Kumada}},\ }\bibfield  {title} {\bibinfo {title} {Spatial structure of the flow through an axisymmetric sudden expansion},\ }\href@noop {} {\bibfield  {journal} {\bibinfo  {journal} {Experiments in Fluids}\ }\textbf {\bibinfo {volume} {34}},\ \bibinfo {pages} {643} (\bibinfo {year} {2003})}\BibitemShut {NoStop}%
\bibitem [{\citenamefont {Teyssandiert}\ and\ \citenamefont {Wilson}(1974)}]{teyssandiert1974analysis}%
  \BibitemOpen
  \bibfield  {author} {\bibinfo {author} {\bibfnamefont {R.}~\bibnamefont {Teyssandiert}}\ and\ \bibinfo {author} {\bibfnamefont {M.}~\bibnamefont {Wilson}},\ }\bibfield  {title} {\bibinfo {title} {An analysis of flow through sudden enlargements in pipes},\ }\href@noop {} {\bibfield  {journal} {\bibinfo  {journal} {Journal of Fluid Mechanics}\ }\textbf {\bibinfo {volume} {64}},\ \bibinfo {pages} {85} (\bibinfo {year} {1974})}\BibitemShut {NoStop}%
\bibitem [{\citenamefont {Cantwell}\ \emph {et~al.}(2010)\citenamefont {Cantwell}, \citenamefont {Barkley},\ and\ \citenamefont {Blackburn}}]{cantwell2010transient}%
  \BibitemOpen
  \bibfield  {author} {\bibinfo {author} {\bibfnamefont {C.}~\bibnamefont {Cantwell}}, \bibinfo {author} {\bibfnamefont {D.}~\bibnamefont {Barkley}},\ and\ \bibinfo {author} {\bibfnamefont {H.}~\bibnamefont {Blackburn}},\ }\bibfield  {title} {\bibinfo {title} {Transient growth analysis of flow through a sudden expansion in a circular pipe},\ }\href@noop {} {\bibfield  {journal} {\bibinfo  {journal} {Physics of Fluids}\ }\textbf {\bibinfo {volume} {22}} (\bibinfo {year} {2010})}\BibitemShut {NoStop}%
\bibitem [{\citenamefont {Javadi}\ and\ \citenamefont {Nilsson}(2015)}]{javadi2015and}%
  \BibitemOpen
  \bibfield  {author} {\bibinfo {author} {\bibfnamefont {A.}~\bibnamefont {Javadi}}\ and\ \bibinfo {author} {\bibfnamefont {H.}~\bibnamefont {Nilsson}},\ }\bibfield  {title} {\bibinfo {title} {Les and des of strongly swirling turbulent flow through a suddenly expanding circular pipe},\ }\href@noop {} {\bibfield  {journal} {\bibinfo  {journal} {Computers \& Fluids}\ }\textbf {\bibinfo {volume} {107}},\ \bibinfo {pages} {301} (\bibinfo {year} {2015})}\BibitemShut {NoStop}%
\bibitem [{\citenamefont {Selvam}\ \emph {et~al.}(2015)\citenamefont {Selvam}, \citenamefont {Peixinho},\ and\ \citenamefont {Willis}}]{selvam2015localised}%
  \BibitemOpen
  \bibfield  {author} {\bibinfo {author} {\bibfnamefont {K.}~\bibnamefont {Selvam}}, \bibinfo {author} {\bibfnamefont {J.}~\bibnamefont {Peixinho}},\ and\ \bibinfo {author} {\bibfnamefont {A.~P.}\ \bibnamefont {Willis}},\ }\bibfield  {title} {\bibinfo {title} {Localised turbulence in a circular pipe flow with gradual expansion},\ }\href@noop {} {\bibfield  {journal} {\bibinfo  {journal} {Journal of Fluid Mechanics}\ }\textbf {\bibinfo {volume} {771}},\ \bibinfo {pages} {R2} (\bibinfo {year} {2015})}\BibitemShut {NoStop}%
\bibitem [{\citenamefont {Lebon}\ \emph {et~al.}(2018)\citenamefont {Lebon}, \citenamefont {Nguyen}, \citenamefont {Peixinho}, \citenamefont {Shadloo},\ and\ \citenamefont {Hadjadj}}]{lebon2018new}%
  \BibitemOpen
  \bibfield  {author} {\bibinfo {author} {\bibfnamefont {B.}~\bibnamefont {Lebon}}, \bibinfo {author} {\bibfnamefont {M.~Q.}\ \bibnamefont {Nguyen}}, \bibinfo {author} {\bibfnamefont {J.}~\bibnamefont {Peixinho}}, \bibinfo {author} {\bibfnamefont {M.~S.}\ \bibnamefont {Shadloo}},\ and\ \bibinfo {author} {\bibfnamefont {A.}~\bibnamefont {Hadjadj}},\ }\bibfield  {title} {\bibinfo {title} {A new mechanism for periodic bursting of the recirculation region in the flow through a sudden expansion in a circular pipe},\ }\href@noop {} {\bibfield  {journal} {\bibinfo  {journal} {Physics of Fluids}\ }\textbf {\bibinfo {volume} {30}} (\bibinfo {year} {2018})}\BibitemShut {NoStop}%
\bibitem [{\citenamefont {Bradshaw}\ and\ \citenamefont {Wong}(1972)}]{bradshaw1972reattachment}%
  \BibitemOpen
  \bibfield  {author} {\bibinfo {author} {\bibfnamefont {P.}~\bibnamefont {Bradshaw}}\ and\ \bibinfo {author} {\bibfnamefont {F.}~\bibnamefont {Wong}},\ }\bibfield  {title} {\bibinfo {title} {The reattachment and relaxation of a turbulent shear layer},\ }\href@noop {} {\bibfield  {journal} {\bibinfo  {journal} {Journal of Fluid Mechanics}\ }\textbf {\bibinfo {volume} {52}},\ \bibinfo {pages} {113} (\bibinfo {year} {1972})}\BibitemShut {NoStop}%
\bibitem [{\citenamefont {Chen}\ \emph {et~al.}(2018)\citenamefont {Chen}, \citenamefont {Asai}, \citenamefont {Nonomura}, \citenamefont {Xi},\ and\ \citenamefont {Liu}}]{chen2018review}%
  \BibitemOpen
  \bibfield  {author} {\bibinfo {author} {\bibfnamefont {L.}~\bibnamefont {Chen}}, \bibinfo {author} {\bibfnamefont {K.}~\bibnamefont {Asai}}, \bibinfo {author} {\bibfnamefont {T.}~\bibnamefont {Nonomura}}, \bibinfo {author} {\bibfnamefont {G.}~\bibnamefont {Xi}},\ and\ \bibinfo {author} {\bibfnamefont {T.}~\bibnamefont {Liu}},\ }\bibfield  {title} {\bibinfo {title} {A review of backward-facing step (bfs) flow mechanisms, heat transfer and control},\ }\href@noop {} {\bibfield  {journal} {\bibinfo  {journal} {Thermal Science and Engineering Progress}\ }\textbf {\bibinfo {volume} {6}},\ \bibinfo {pages} {194} (\bibinfo {year} {2018})}\BibitemShut {NoStop}%
\bibitem [{\citenamefont {Scarano}\ \emph {et~al.}(1999)\citenamefont {Scarano}, \citenamefont {Benocci},\ and\ \citenamefont {Riethmuller}}]{scarano1999pattern}%
  \BibitemOpen
  \bibfield  {author} {\bibinfo {author} {\bibfnamefont {F.}~\bibnamefont {Scarano}}, \bibinfo {author} {\bibfnamefont {C.}~\bibnamefont {Benocci}},\ and\ \bibinfo {author} {\bibfnamefont {M.}~\bibnamefont {Riethmuller}},\ }\bibfield  {title} {\bibinfo {title} {Pattern recognition analysis of the turbulent flow past a backward facing step},\ }\href@noop {} {\bibfield  {journal} {\bibinfo  {journal} {Physics of Fluids}\ }\textbf {\bibinfo {volume} {11}},\ \bibinfo {pages} {3808} (\bibinfo {year} {1999})}\BibitemShut {NoStop}%
\bibitem [{\citenamefont {Bernal}\ and\ \citenamefont {Roshko}(1986)}]{bernal1986streamwise}%
  \BibitemOpen
  \bibfield  {author} {\bibinfo {author} {\bibfnamefont {L.}~\bibnamefont {Bernal}}\ and\ \bibinfo {author} {\bibfnamefont {A.}~\bibnamefont {Roshko}},\ }\bibfield  {title} {\bibinfo {title} {Streamwise vortex structure in plane mixing layers},\ }\href@noop {} {\bibfield  {journal} {\bibinfo  {journal} {Journal of Fluid Mechanics}\ }\textbf {\bibinfo {volume} {170}},\ \bibinfo {pages} {499} (\bibinfo {year} {1986})}\BibitemShut {NoStop}%
\bibitem [{\citenamefont {Bell}\ and\ \citenamefont {Mehta}(1992)}]{bell1992measurements}%
  \BibitemOpen
  \bibfield  {author} {\bibinfo {author} {\bibfnamefont {J.~H.}\ \bibnamefont {Bell}}\ and\ \bibinfo {author} {\bibfnamefont {R.~D.}\ \bibnamefont {Mehta}},\ }\bibfield  {title} {\bibinfo {title} {Measurements of the streamwise vortical structures in a plane mixing layer},\ }\href@noop {} {\bibfield  {journal} {\bibinfo  {journal} {Journal of Fluid Mechanics}\ }\textbf {\bibinfo {volume} {239}},\ \bibinfo {pages} {213} (\bibinfo {year} {1992})}\BibitemShut {NoStop}%
\bibitem [{\citenamefont {Agarwal}\ \emph {et~al.}(2023)\citenamefont {Agarwal}, \citenamefont {Ram}, \citenamefont {Lu},\ and\ \citenamefont {Katz}}]{agarwal2023pressure}%
  \BibitemOpen
  \bibfield  {author} {\bibinfo {author} {\bibfnamefont {K.}~\bibnamefont {Agarwal}}, \bibinfo {author} {\bibfnamefont {O.}~\bibnamefont {Ram}}, \bibinfo {author} {\bibfnamefont {Y.}~\bibnamefont {Lu}},\ and\ \bibinfo {author} {\bibfnamefont {J.}~\bibnamefont {Katz}},\ }\bibfield  {title} {\bibinfo {title} {On the pressure field, nuclei dynamics and their relation to cavitation inception in a turbulent shear layer},\ }\href@noop {} {\bibfield  {journal} {\bibinfo  {journal} {Journal of Fluid Mechanics}\ }\textbf {\bibinfo {volume} {966}},\ \bibinfo {pages} {A31} (\bibinfo {year} {2023})}\BibitemShut {NoStop}%
\bibitem [{\citenamefont {Kim}\ \emph {et~al.}(1980)\citenamefont {Kim}, \citenamefont {Kline},\ and\ \citenamefont {Johnston}}]{kim1978investigation}%
  \BibitemOpen
  \bibfield  {author} {\bibinfo {author} {\bibfnamefont {J.}~\bibnamefont {Kim}}, \bibinfo {author} {\bibfnamefont {S.}~\bibnamefont {Kline}},\ and\ \bibinfo {author} {\bibfnamefont {J.}~\bibnamefont {Johnston}},\ }\bibfield  {title} {\bibinfo {title} {Investigation of separation and reattachment of a turbulent shear layer: flow over a backward-facing step.},\ }\href@noop {} {\bibfield  {journal} {\bibinfo  {journal} {Journal of Fluids Engineering}\ }\textbf {\bibinfo {volume} {102}},\ \bibinfo {pages} {302} (\bibinfo {year} {1980})}\BibitemShut {NoStop}%
\bibitem [{\citenamefont {Kostas}\ \emph {et~al.}(2002)\citenamefont {Kostas}, \citenamefont {Soria},\ and\ \citenamefont {Chong}}]{kostas2002particle}%
  \BibitemOpen
  \bibfield  {author} {\bibinfo {author} {\bibfnamefont {J.}~\bibnamefont {Kostas}}, \bibinfo {author} {\bibfnamefont {J.}~\bibnamefont {Soria}},\ and\ \bibinfo {author} {\bibfnamefont {M.}~\bibnamefont {Chong}},\ }\bibfield  {title} {\bibinfo {title} {Particle image velocimetry measurements of a backward-facing step flow},\ }\href@noop {} {\bibfield  {journal} {\bibinfo  {journal} {Experiments in fluids}\ }\textbf {\bibinfo {volume} {33}},\ \bibinfo {pages} {838} (\bibinfo {year} {2002})}\BibitemShut {NoStop}%
\bibitem [{\citenamefont {Bhattacharjee}\ \emph {et~al.}(1986)\citenamefont {Bhattacharjee}, \citenamefont {Scheelke},\ and\ \citenamefont {Troutt}}]{bhattacharjee1986modification}%
  \BibitemOpen
  \bibfield  {author} {\bibinfo {author} {\bibfnamefont {S.}~\bibnamefont {Bhattacharjee}}, \bibinfo {author} {\bibfnamefont {B.}~\bibnamefont {Scheelke}},\ and\ \bibinfo {author} {\bibfnamefont {T.}~\bibnamefont {Troutt}},\ }\bibfield  {title} {\bibinfo {title} {Modification of vortex interactions in a reattaching separated flow},\ }\href@noop {} {\bibfield  {journal} {\bibinfo  {journal} {AIAA journal}\ }\textbf {\bibinfo {volume} {24}},\ \bibinfo {pages} {623} (\bibinfo {year} {1986})}\BibitemShut {NoStop}%
\bibitem [{\citenamefont {Ma}\ and\ \citenamefont {Schr{\"o}der}(2017)}]{ma2017analysis}%
  \BibitemOpen
  \bibfield  {author} {\bibinfo {author} {\bibfnamefont {X.}~\bibnamefont {Ma}}\ and\ \bibinfo {author} {\bibfnamefont {A.}~\bibnamefont {Schr{\"o}der}},\ }\bibfield  {title} {\bibinfo {title} {Analysis of flapping motion of reattaching shear layer behind a two-dimensional backward-facing step},\ }\href@noop {} {\bibfield  {journal} {\bibinfo  {journal} {Physics of Fluids}\ }\textbf {\bibinfo {volume} {29}} (\bibinfo {year} {2017})}\BibitemShut {NoStop}%
\bibitem [{\citenamefont {Spazzini}\ \emph {et~al.}(2001)\citenamefont {Spazzini}, \citenamefont {Iuso}, \citenamefont {Onorato}, \citenamefont {Zurlo},\ and\ \citenamefont {Di~C.}}]{spazzini2001unsteady}%
  \BibitemOpen
  \bibfield  {author} {\bibinfo {author} {\bibfnamefont {P.~G.}\ \bibnamefont {Spazzini}}, \bibinfo {author} {\bibfnamefont {G.}~\bibnamefont {Iuso}}, \bibinfo {author} {\bibfnamefont {M.}~\bibnamefont {Onorato}}, \bibinfo {author} {\bibfnamefont {N.}~\bibnamefont {Zurlo}},\ and\ \bibinfo {author} {\bibfnamefont {G.~M.}\ \bibnamefont {Di~C.}},\ }\bibfield  {title} {\bibinfo {title} {Unsteady behavior of back-facing step flow},\ }\href@noop {} {\bibfield  {journal} {\bibinfo  {journal} {Experiments in fluids}\ }\textbf {\bibinfo {volume} {30}},\ \bibinfo {pages} {551} (\bibinfo {year} {2001})}\BibitemShut {NoStop}%
\bibitem [{\citenamefont {Sampath}\ and\ \citenamefont {Chakravarthy}(2014)}]{sampath2014proper}%
  \BibitemOpen
  \bibfield  {author} {\bibinfo {author} {\bibfnamefont {R.}~\bibnamefont {Sampath}}\ and\ \bibinfo {author} {\bibfnamefont {S.}~\bibnamefont {Chakravarthy}},\ }\bibfield  {title} {\bibinfo {title} {Proper orthogonal and dynamic mode decompositions of time-resolved piv of confined backward-facing step flow},\ }\href@noop {} {\bibfield  {journal} {\bibinfo  {journal} {Experiments in fluids}\ }\textbf {\bibinfo {volume} {55}},\ \bibinfo {pages} {1} (\bibinfo {year} {2014})}\BibitemShut {NoStop}%
\bibitem [{\citenamefont {Ma}\ \emph {et~al.}(2022)\citenamefont {Ma}, \citenamefont {Tang},\ and\ \citenamefont {Jiang}}]{ma2022investigation}%
  \BibitemOpen
  \bibfield  {author} {\bibinfo {author} {\bibfnamefont {X.}~\bibnamefont {Ma}}, \bibinfo {author} {\bibfnamefont {Z.}~\bibnamefont {Tang}},\ and\ \bibinfo {author} {\bibfnamefont {N.}~\bibnamefont {Jiang}},\ }\bibfield  {title} {\bibinfo {title} {Investigation of spanwise coherent structures in turbulent backward-facing step flow by time-resolved piv},\ }\href@noop {} {\bibfield  {journal} {\bibinfo  {journal} {Experimental Thermal and Fluid Science}\ }\textbf {\bibinfo {volume} {132}},\ \bibinfo {pages} {110569} (\bibinfo {year} {2022})}\BibitemShut {NoStop}%
\bibitem [{\citenamefont {Sch{\"a}fer}\ \emph {et~al.}(2009)\citenamefont {Sch{\"a}fer}, \citenamefont {Breuer},\ and\ \citenamefont {Durst}}]{schafer2009dynamics}%
  \BibitemOpen
  \bibfield  {author} {\bibinfo {author} {\bibfnamefont {F.}~\bibnamefont {Sch{\"a}fer}}, \bibinfo {author} {\bibfnamefont {M.}~\bibnamefont {Breuer}},\ and\ \bibinfo {author} {\bibfnamefont {F.}~\bibnamefont {Durst}},\ }\bibfield  {title} {\bibinfo {title} {The dynamics of the transitional flow over a backward-facing step},\ }\href@noop {} {\bibfield  {journal} {\bibinfo  {journal} {Journal of Fluid Mechanics}\ }\textbf {\bibinfo {volume} {623}},\ \bibinfo {pages} {85} (\bibinfo {year} {2009})}\BibitemShut {NoStop}%
\bibitem [{\citenamefont {Browand}(1966)}]{browand1966experimental}%
  \BibitemOpen
  \bibfield  {author} {\bibinfo {author} {\bibfnamefont {F.~K.}\ \bibnamefont {Browand}},\ }\bibfield  {title} {\bibinfo {title} {An experimental investigation of the instability of an incompressible, separated shear layer},\ }\href@noop {} {\bibfield  {journal} {\bibinfo  {journal} {Journal of Fluid Mechanics}\ }\textbf {\bibinfo {volume} {26}},\ \bibinfo {pages} {281} (\bibinfo {year} {1966})}\BibitemShut {NoStop}%
\bibitem [{\citenamefont {Winant}\ and\ \citenamefont {Browand}(1974)}]{winant1974vortex}%
  \BibitemOpen
  \bibfield  {author} {\bibinfo {author} {\bibfnamefont {C.~D.}\ \bibnamefont {Winant}}\ and\ \bibinfo {author} {\bibfnamefont {F.~K.}\ \bibnamefont {Browand}},\ }\bibfield  {title} {\bibinfo {title} {Vortex pairing: the mechanism of turbulent mixing-layer growth at moderate reynolds number},\ }\href@noop {} {\bibfield  {journal} {\bibinfo  {journal} {Journal of Fluid Mechanics}\ }\textbf {\bibinfo {volume} {63}},\ \bibinfo {pages} {237} (\bibinfo {year} {1974})}\BibitemShut {NoStop}%
\bibitem [{\citenamefont {Troutt}\ \emph {et~al.}(1984)\citenamefont {Troutt}, \citenamefont {Scheelke},\ and\ \citenamefont {Norman}}]{troutt1984organized}%
  \BibitemOpen
  \bibfield  {author} {\bibinfo {author} {\bibfnamefont {T.}~\bibnamefont {Troutt}}, \bibinfo {author} {\bibfnamefont {B.}~\bibnamefont {Scheelke}},\ and\ \bibinfo {author} {\bibfnamefont {T.}~\bibnamefont {Norman}},\ }\bibfield  {title} {\bibinfo {title} {Organized structures in a reattaching separated flow field},\ }\href@noop {} {\bibfield  {journal} {\bibinfo  {journal} {Journal of Fluid Mechanics}\ }\textbf {\bibinfo {volume} {143}},\ \bibinfo {pages} {413} (\bibinfo {year} {1984})}\BibitemShut {NoStop}%
\bibitem [{\citenamefont {Kasagi}\ and\ \citenamefont {Matsunaga}(1995)}]{kasagi1995three}%
  \BibitemOpen
  \bibfield  {author} {\bibinfo {author} {\bibfnamefont {N.}~\bibnamefont {Kasagi}}\ and\ \bibinfo {author} {\bibfnamefont {A.}~\bibnamefont {Matsunaga}},\ }\bibfield  {title} {\bibinfo {title} {Three-dimensional particle-tracking velocimetry measurement of turbulence statistics and energy budget in a backward-facing step flow},\ }\href@noop {} {\bibfield  {journal} {\bibinfo  {journal} {International journal of heat and fluid flow}\ }\textbf {\bibinfo {volume} {16}},\ \bibinfo {pages} {477} (\bibinfo {year} {1995})}\BibitemShut {NoStop}%
\bibitem [{\citenamefont {Fessler}\ and\ \citenamefont {Eaton}(1999)}]{fessler1999turbulence}%
  \BibitemOpen
  \bibfield  {author} {\bibinfo {author} {\bibfnamefont {J.~R.}\ \bibnamefont {Fessler}}\ and\ \bibinfo {author} {\bibfnamefont {J.~K.}\ \bibnamefont {Eaton}},\ }\bibfield  {title} {\bibinfo {title} {Turbulence modification by particles in a backward-facing step flow},\ }\href@noop {} {\bibfield  {journal} {\bibinfo  {journal} {Journal of Fluid Mechanics}\ }\textbf {\bibinfo {volume} {394}},\ \bibinfo {pages} {97} (\bibinfo {year} {1999})}\BibitemShut {NoStop}%
\bibitem [{\citenamefont {Piirto}\ \emph {et~al.}(2003)\citenamefont {Piirto}, \citenamefont {Saarenrinne}, \citenamefont {Eloranta},\ and\ \citenamefont {Karvinen}}]{piirto2003measuring}%
  \BibitemOpen
  \bibfield  {author} {\bibinfo {author} {\bibfnamefont {M.}~\bibnamefont {Piirto}}, \bibinfo {author} {\bibfnamefont {P.}~\bibnamefont {Saarenrinne}}, \bibinfo {author} {\bibfnamefont {H.}~\bibnamefont {Eloranta}},\ and\ \bibinfo {author} {\bibfnamefont {R.}~\bibnamefont {Karvinen}},\ }\bibfield  {title} {\bibinfo {title} {Measuring turbulence energy with piv in a backward-facing step flow},\ }\href@noop {} {\bibfield  {journal} {\bibinfo  {journal} {Experiments in fluids}\ }\textbf {\bibinfo {volume} {35}},\ \bibinfo {pages} {219} (\bibinfo {year} {2003})}\BibitemShut {NoStop}%
\bibitem [{\citenamefont {Hussain}(1986)}]{hussain1986coherent}%
  \BibitemOpen
  \bibfield  {author} {\bibinfo {author} {\bibfnamefont {A.~F.}\ \bibnamefont {Hussain}},\ }\bibfield  {title} {\bibinfo {title} {Coherent structures and turbulence},\ }\href@noop {} {\bibfield  {journal} {\bibinfo  {journal} {Journal of Fluid Mechanics}\ }\textbf {\bibinfo {volume} {173}},\ \bibinfo {pages} {303} (\bibinfo {year} {1986})}\BibitemShut {NoStop}%
\bibitem [{\citenamefont {Lee}\ \emph {et~al.}(2004)\citenamefont {Lee}, \citenamefont {Ahn},\ and\ \citenamefont {Sung}}]{lee2004three}%
  \BibitemOpen
  \bibfield  {author} {\bibinfo {author} {\bibfnamefont {I.}~\bibnamefont {Lee}}, \bibinfo {author} {\bibfnamefont {S.}~\bibnamefont {Ahn}},\ and\ \bibinfo {author} {\bibfnamefont {H.~J.}\ \bibnamefont {Sung}},\ }\bibfield  {title} {\bibinfo {title} {Three-dimensional coherent structure in a separated and reattaching flow over a backward-facing step},\ }\href@noop {} {\bibfield  {journal} {\bibinfo  {journal} {Experiments in fluids}\ }\textbf {\bibinfo {volume} {36}},\ \bibinfo {pages} {373} (\bibinfo {year} {2004})}\BibitemShut {NoStop}%
\bibitem [{\citenamefont {Wang}\ \emph {et~al.}(2019)\citenamefont {Wang}, \citenamefont {Gao}, \citenamefont {Wu}, \citenamefont {Zhu}, \citenamefont {Dai},\ and\ \citenamefont {Liao}}]{wang2019experimental}%
  \BibitemOpen
  \bibfield  {author} {\bibinfo {author} {\bibfnamefont {F.}~\bibnamefont {Wang}}, \bibinfo {author} {\bibfnamefont {A.}~\bibnamefont {Gao}}, \bibinfo {author} {\bibfnamefont {S.}~\bibnamefont {Wu}}, \bibinfo {author} {\bibfnamefont {S.}~\bibnamefont {Zhu}}, \bibinfo {author} {\bibfnamefont {J.}~\bibnamefont {Dai}},\ and\ \bibinfo {author} {\bibfnamefont {Q.}~\bibnamefont {Liao}},\ }\bibfield  {title} {\bibinfo {title} {Experimental investigation of coherent vortex structures in a backward-facing step flow},\ }\href@noop {} {\bibfield  {journal} {\bibinfo  {journal} {Water}\ }\textbf {\bibinfo {volume} {11}},\ \bibinfo {pages} {2629} (\bibinfo {year} {2019})}\BibitemShut {NoStop}%
\bibitem [{\citenamefont {Ahmadpour}\ \emph {et~al.}(2016)\citenamefont {Ahmadpour}, \citenamefont {Abadi}, \citenamefont {R.},\ and\ \citenamefont {Kouhikamali}}]{ahmadpour2016numerical}%
  \BibitemOpen
  \bibfield  {author} {\bibinfo {author} {\bibfnamefont {A.}~\bibnamefont {Ahmadpour}}, \bibinfo {author} {\bibfnamefont {S.}~\bibnamefont {Abadi}}, \bibinfo {author} {\bibfnamefont {N.}~\bibnamefont {R.}},\ and\ \bibinfo {author} {\bibfnamefont {R.}~\bibnamefont {Kouhikamali}},\ }\bibfield  {title} {\bibinfo {title} {Numerical simulation of two-phase gas--liquid flow through gradual expansions/contractions},\ }\href@noop {} {\bibfield  {journal} {\bibinfo  {journal} {International Journal of Multiphase Flow}\ }\textbf {\bibinfo {volume} {79}},\ \bibinfo {pages} {31} (\bibinfo {year} {2016})}\BibitemShut {NoStop}%
\bibitem [{\citenamefont {Choi}\ \emph {et~al.}(2016)\citenamefont {Choi}, \citenamefont {Nguyen},\ and\ \citenamefont {Nguyen}}]{choi2016numerical}%
  \BibitemOpen
  \bibfield  {author} {\bibinfo {author} {\bibfnamefont {H.}~\bibnamefont {Choi}}, \bibinfo {author} {\bibfnamefont {V.~T.}\ \bibnamefont {Nguyen}},\ and\ \bibinfo {author} {\bibfnamefont {J.}~\bibnamefont {Nguyen}},\ }\bibfield  {title} {\bibinfo {title} {Numerical investigation of backward facing step flow over various step angles},\ }\href@noop {} {\bibfield  {journal} {\bibinfo  {journal} {Procedia Engineering}\ }\textbf {\bibinfo {volume} {154}},\ \bibinfo {pages} {420} (\bibinfo {year} {2016})}\BibitemShut {NoStop}%
\bibitem [{\citenamefont {Danane}\ \emph {et~al.}(2020)\citenamefont {Danane}, \citenamefont {Boudiaf}, \citenamefont {Mahfoud}, \citenamefont {Ouyahia}, \citenamefont {Labsi},\ and\ \citenamefont {Benkahla}}]{danane2020effect}%
  \BibitemOpen
  \bibfield  {author} {\bibinfo {author} {\bibfnamefont {F.}~\bibnamefont {Danane}}, \bibinfo {author} {\bibfnamefont {A.}~\bibnamefont {Boudiaf}}, \bibinfo {author} {\bibfnamefont {O.}~\bibnamefont {Mahfoud}}, \bibinfo {author} {\bibfnamefont {S.}~\bibnamefont {Ouyahia}}, \bibinfo {author} {\bibfnamefont {N.}~\bibnamefont {Labsi}},\ and\ \bibinfo {author} {\bibfnamefont {Y.~K.}\ \bibnamefont {Benkahla}},\ }\bibfield  {title} {\bibinfo {title} {Effect of backward facing step shape on 3d mixed convection of bingham fluid},\ }\href@noop {} {\bibfield  {journal} {\bibinfo  {journal} {International journal of thermal sciences}\ }\textbf {\bibinfo {volume} {147}},\ \bibinfo {pages} {106116} (\bibinfo {year} {2020})}\BibitemShut {NoStop}%
\bibitem [{\citenamefont {Bai}\ and\ \citenamefont {Katz}(2014)}]{bai2014refractive}%
  \BibitemOpen
  \bibfield  {author} {\bibinfo {author} {\bibfnamefont {K.}~\bibnamefont {Bai}}\ and\ \bibinfo {author} {\bibfnamefont {J.}~\bibnamefont {Katz}},\ }\bibfield  {title} {\bibinfo {title} {On the refractive index of sodium iodide solutions for index matching in piv},\ }\href@noop {} {\bibfield  {journal} {\bibinfo  {journal} {Experiments in fluids}\ }\textbf {\bibinfo {volume} {55}},\ \bibinfo {pages} {1} (\bibinfo {year} {2014})}\BibitemShut {NoStop}%
\bibitem [{\citenamefont {Meinhart}\ \emph {et~al.}(2000)\citenamefont {Meinhart}, \citenamefont {Wereley},\ and\ \citenamefont {Santiago}}]{meinhart2000piv}%
  \BibitemOpen
  \bibfield  {author} {\bibinfo {author} {\bibfnamefont {C.~D.}\ \bibnamefont {Meinhart}}, \bibinfo {author} {\bibfnamefont {S.~T.}\ \bibnamefont {Wereley}},\ and\ \bibinfo {author} {\bibfnamefont {J.~G.}\ \bibnamefont {Santiago}},\ }\bibfield  {title} {\bibinfo {title} {A piv algorithm for estimating time-averaged velocity fields},\ }\href@noop {} {\bibfield  {journal} {\bibinfo  {journal} {Journal of fluids engineering}\ }\textbf {\bibinfo {volume} {122}},\ \bibinfo {pages} {285} (\bibinfo {year} {2000})}\BibitemShut {NoStop}%
\bibitem [{\citenamefont {Saarenrinne}\ \emph {et~al.}(2001)\citenamefont {Saarenrinne}, \citenamefont {Piirto},\ and\ \citenamefont {Eloranta}}]{saarenrinne2001experiences}%
  \BibitemOpen
  \bibfield  {author} {\bibinfo {author} {\bibfnamefont {P.}~\bibnamefont {Saarenrinne}}, \bibinfo {author} {\bibfnamefont {M.}~\bibnamefont {Piirto}},\ and\ \bibinfo {author} {\bibfnamefont {H.}~\bibnamefont {Eloranta}},\ }\bibfield  {title} {\bibinfo {title} {Experiences of turbulence measurement with piv},\ }\href {https://doi.org/10.1007/s003480000179} {\bibfield  {journal} {\bibinfo  {journal} {Experiments in Fluids}\ }\textbf {\bibinfo {volume} {30}},\ \bibinfo {pages} {282} (\bibinfo {year} {2001})}\BibitemShut {NoStop}%
\bibitem [{\citenamefont {Brown}\ and\ \citenamefont {Roshko}(1974)}]{BrownRoshko1974}%
  \BibitemOpen
  \bibfield  {author} {\bibinfo {author} {\bibfnamefont {G.~L.}\ \bibnamefont {Brown}}\ and\ \bibinfo {author} {\bibfnamefont {A.}~\bibnamefont {Roshko}},\ }\bibfield  {title} {\bibinfo {title} {On density effects and large structure in turbulent mixing layers},\ }\href {https://doi.org/10.1017/S0022112074002034} {\bibfield  {journal} {\bibinfo  {journal} {Journal of Fluid Mechanics}\ }\textbf {\bibinfo {volume} {64}},\ \bibinfo {pages} {775} (\bibinfo {year} {1974})}\BibitemShut {NoStop}%
\bibitem [{\citenamefont {Tinney}\ \emph {et~al.}(2006)\citenamefont {Tinney}, \citenamefont {Glauser}, \citenamefont {Eaton},\ and\ \citenamefont {Taylor}}]{tinney2006low}%
  \BibitemOpen
  \bibfield  {author} {\bibinfo {author} {\bibfnamefont {C.}~\bibnamefont {Tinney}}, \bibinfo {author} {\bibfnamefont {M.}~\bibnamefont {Glauser}}, \bibinfo {author} {\bibfnamefont {E.}~\bibnamefont {Eaton}},\ and\ \bibinfo {author} {\bibfnamefont {J.}~\bibnamefont {Taylor}},\ }\bibfield  {title} {\bibinfo {title} {Low-dimensional azimuthal characteristics of suddenly expanding axisymmetric flows},\ }\href@noop {} {\bibfield  {journal} {\bibinfo  {journal} {Journal of Fluid Mechanics}\ }\textbf {\bibinfo {volume} {567}},\ \bibinfo {pages} {141} (\bibinfo {year} {2006})}\BibitemShut {NoStop}%
\bibitem [{\citenamefont {Jovic}(1996)}]{jovic1996experimental}%
  \BibitemOpen
  \bibfield  {author} {\bibinfo {author} {\bibfnamefont {S.}~\bibnamefont {Jovic}},\ }\href@noop {} {\emph {\bibinfo {title} {An Experimental Study of a Separated/Reattached Flow Behind a Backward-Facing Step. Re (sub h)= 37,000}}},\ \bibinfo {type} {Tech. Rep.}\ \bibinfo {number} {110384}\ (\bibinfo  {institution} {NASA Ames Research Center},\ \bibinfo {year} {1996})\BibitemShut {NoStop}%
\bibitem [{\citenamefont {Choi}\ and\ \citenamefont {Lumley}(2001)}]{choi2001return}%
  \BibitemOpen
  \bibfield  {author} {\bibinfo {author} {\bibfnamefont {K.}~\bibnamefont {Choi}}\ and\ \bibinfo {author} {\bibfnamefont {J.~L.}\ \bibnamefont {Lumley}},\ }\bibfield  {title} {\bibinfo {title} {The return to isotropy of homogeneous turbulence},\ }\href@noop {} {\bibfield  {journal} {\bibinfo  {journal} {Journal of Fluid Mechanics}\ }\textbf {\bibinfo {volume} {436}},\ \bibinfo {pages} {59} (\bibinfo {year} {2001})}\BibitemShut {NoStop}%
\bibitem [{\citenamefont {Dey}\ \emph {et~al.}(2020)\citenamefont {Dey}, \citenamefont {Paul}, \citenamefont {Ali},\ and\ \citenamefont {Padhi}}]{dey2020reynolds}%
  \BibitemOpen
  \bibfield  {author} {\bibinfo {author} {\bibfnamefont {S.}~\bibnamefont {Dey}}, \bibinfo {author} {\bibfnamefont {P.}~\bibnamefont {Paul}}, \bibinfo {author} {\bibfnamefont {S.}~\bibnamefont {Ali}},\ and\ \bibinfo {author} {\bibfnamefont {E.}~\bibnamefont {Padhi}},\ }\bibfield  {title} {\bibinfo {title} {Reynolds stress anisotropy in flow over two-dimensional rigid dunes},\ }\href@noop {} {\bibfield  {journal} {\bibinfo  {journal} {Proceedings of the Royal Society A}\ }\textbf {\bibinfo {volume} {476}},\ \bibinfo {pages} {20200638} (\bibinfo {year} {2020})}\BibitemShut {NoStop}%
\bibitem [{\citenamefont {Fang}\ \emph {et~al.}(2021)\citenamefont {Fang}, \citenamefont {Tachie},\ and\ \citenamefont {Bergstrom}}]{fang2021direct}%
  \BibitemOpen
  \bibfield  {author} {\bibinfo {author} {\bibfnamefont {X.}~\bibnamefont {Fang}}, \bibinfo {author} {\bibfnamefont {M.~F.}\ \bibnamefont {Tachie}},\ and\ \bibinfo {author} {\bibfnamefont {D.~J.}\ \bibnamefont {Bergstrom}},\ }\bibfield  {title} {\bibinfo {title} {Direct numerical simulation of turbulent flow separation induced by a forward-facing step},\ }\href@noop {} {\bibfield  {journal} {\bibinfo  {journal} {International Journal of Heat and Fluid Flow}\ }\textbf {\bibinfo {volume} {87}},\ \bibinfo {pages} {108753} (\bibinfo {year} {2021})}\BibitemShut {NoStop}%
\bibitem [{\citenamefont {Jang}\ \emph {et~al.}(2011)\citenamefont {Jang}, \citenamefont {Sung},\ and\ \citenamefont {Krogstad}}]{jang2011effects}%
  \BibitemOpen
  \bibfield  {author} {\bibinfo {author} {\bibfnamefont {S.~J.}\ \bibnamefont {Jang}}, \bibinfo {author} {\bibfnamefont {H.~J.}\ \bibnamefont {Sung}},\ and\ \bibinfo {author} {\bibfnamefont {P.-{\AA}.}\ \bibnamefont {Krogstad}},\ }\bibfield  {title} {\bibinfo {title} {Effects of an axisymmetric contraction on a turbulent pipe flow},\ }\href@noop {} {\bibfield  {journal} {\bibinfo  {journal} {Journal of fluid mechanics}\ }\textbf {\bibinfo {volume} {687}},\ \bibinfo {pages} {376} (\bibinfo {year} {2011})}\BibitemShut {NoStop}%
\bibitem [{\citenamefont {Emory}\ and\ \citenamefont {Iaccarino}(2014)}]{emory2014visualizing}%
  \BibitemOpen
  \bibfield  {author} {\bibinfo {author} {\bibfnamefont {M.}~\bibnamefont {Emory}}\ and\ \bibinfo {author} {\bibfnamefont {G.}~\bibnamefont {Iaccarino}},\ }\bibfield  {title} {\bibinfo {title} {Visualizing turbulence anisotropy in the spatial domain with componentality contours},\ }\href {https://ctr.stanford.edu/research/annual-research-briefs} {\bibfield  {journal} {\bibinfo  {journal} {Center for Turbulence Research Annual Research Briefs}\ ,\ \bibinfo {pages} {123}} (\bibinfo {year} {2014})}\BibitemShut {NoStop}%
\bibitem [{\citenamefont {Simonsen}\ and\ \citenamefont {Krogstad}(2005)}]{simonsen2005turbulent}%
  \BibitemOpen
  \bibfield  {author} {\bibinfo {author} {\bibfnamefont {A.}~\bibnamefont {Simonsen}}\ and\ \bibinfo {author} {\bibfnamefont {P.-{\AA}.}\ \bibnamefont {Krogstad}},\ }\bibfield  {title} {\bibinfo {title} {Turbulent stress invariant analysis: Clarification of existing terminology},\ }\href@noop {} {\bibfield  {journal} {\bibinfo  {journal} {Physics of Fluids}\ }\textbf {\bibinfo {volume} {17}} (\bibinfo {year} {2005})}\BibitemShut {NoStop}%
\bibitem [{\citenamefont {Pope}(2000)}]{Pope_2000}%
  \BibitemOpen
  \bibfield  {author} {\bibinfo {author} {\bibfnamefont {S.~B.}\ \bibnamefont {Pope}},\ }\bibinfo {title} {Turbulent flows}\ (\bibinfo  {publisher} {Cambridge University Press},\ \bibinfo {year} {2000})\BibitemShut {NoStop}%
\bibitem [{\citenamefont {Wieneke}(2017)}]{wieneke_piv_2017}%
  \BibitemOpen
  \bibfield  {author} {\bibinfo {author} {\bibfnamefont {B.}~\bibnamefont {Wieneke}},\ }\emph {\bibinfo {title} {{PIV} {Uncertainty} {Quantification} and {Beyond}}},\ \href {https://doi.org/10.4233/uuid:4ca8c0b8-0835-47c3-8523-12fc356768f3} {\bibinfo {type} {Dissertation ({TU} {Delft})}},\ \bibinfo  {school} {Delft University of Technology} (\bibinfo {year} {2017}),\ \bibinfo {note} {iSBN: 9789492516886}\BibitemShut {NoStop}%
\bibitem [{\citenamefont {Sciacchitano}\ and\ \citenamefont {Wieneke}(2016)}]{sciacchitano2016piv}%
  \BibitemOpen
  \bibfield  {author} {\bibinfo {author} {\bibfnamefont {A.}~\bibnamefont {Sciacchitano}}\ and\ \bibinfo {author} {\bibfnamefont {B.}~\bibnamefont {Wieneke}},\ }\bibfield  {title} {\bibinfo {title} {Piv uncertainty propagation},\ }\href@noop {} {\bibfield  {journal} {\bibinfo  {journal} {Measurement Science and Technology}\ }\textbf {\bibinfo {volume} {27}},\ \bibinfo {pages} {084006} (\bibinfo {year} {2016})}\BibitemShut {NoStop}%
\end{thebibliography}%

\end{document}